\documentclass[onecolumn,aps,prd,12pt,superscriptaddress,nofootinbib,floatfix]{revtex4-2}

\usepackage{graphicx}
\graphicspath{{figures/}}
\usepackage[T1]{fontenc}
\usepackage{lmodern}
\usepackage[utf8]{inputenc}
\usepackage{mathtools,amsmath,amssymb,amsfonts,bm}
\usepackage{amsthm}
\usepackage{dcolumn,array,booktabs,multirow,adjustbox}
\usepackage{microtype}
\usepackage{xcolor}
\usepackage[colorlinks,linktocpage,linkcolor=blue,citecolor=blue,urlcolor=blue]{hyperref}
\usepackage[capitalise]{cleveref}

\newcommand{\dd}{\mathop{}\!\mathrm{d}}
\newcommand{\av}[1]{\left\langle #1\right\rangle}

\begin{document}

\title{Pulsar timing response and observational constraints on continuous-spin gravity}

\author{Jing-Hong Han}
\affiliation{Department of Applied Physics, College of Science, China Agricultural University, 17 Qinghua East Road, Haidian District, Beijing 100083, China}

\author{Zhi-Chao Zhao}
\email[Corresponding author: ]{zhaozc@cau.edu.cn}
\affiliation{Department of Applied Physics, College of Science, China Agricultural University, 17 Qinghua East Road, Haidian District, Beijing 100083, China}













\begin{abstract}
Continuous-spin particles form the most general massless integer-helicity irreducible representations of the Poincar\'e group, characterized by a dimensionful continuous-spin scale $\rho_g$ set by a nonzero Pauli--Lubanski invariant. A single continuous-spin representation contains an infinite tower of integer helicities $h\in\mathbb Z$ that mix under Lorentz boosts. Since continuous-spin effects are controlled by $\alpha\equiv\rho_g/(\hbar\omega)$ at gravitational-wave angular frequency $\omega$, nanohertz pulsar timing arrays are naturally sensitive to ultralow values of $\rho_g$. Starting from the gauge-invariant worldline coupling of linearized continuous-spin gravity, we derive the one-way pulsar timing redshift response and the overlap reduction function of an isotropic stochastic background for arbitrary integer helicity. We show that while photon propagation is modulated across all helicities, only the vector modes ($h=\pm1$) perturb the four-velocity of a static observer, and that the tensor modes ($h=\pm2$) smoothly recover the Hellings--Downs correlation of general relativity as $\rho_g/(\hbar\omega)\to 0$. Modeling an unpolarized $h=\pm2$ background in the solar-system barycentric frame and jointly fitting the NANOGrav 15-year data with pulsar red noise, we obtain the $95\%$ upper limits $\rho^{\mathcal{U}}_g<7.03\times10^{-23}\,\mathrm{eV}$ and $\rho^{\log \mathcal{U}}_g<4.79\times10^{-23}\,\mathrm{eV}$ under uniform and log-uniform priors, respectively. Under the uniform prior, we obtain $\gamma_{\mathrm{GW}} = 4.89^{+0.85}_{-0.96}$, which brings the spectral index closer to the canonical value $\gamma_{\rm GW}=13/3$ expected for circular SMBHBs than standard General Relativity.
\end{abstract}

\maketitle 
\section{Introduction}
\label{sec:intro}
Pulsar timing arrays (PTAs) time ensembles of Galactic millisecond pulsars over decade-long baselines to search for nanohertz gravitational waves~\cite{Sazhin:1978myk,Detweiler:1979wn,Foster1990,Jenet:2005pv,Hobbs:2009yy}. In general relativity (GR), an isotropic, unpolarized stochastic gravitational-wave background (SGWB) induces a characteristic quadrupole-dominated inter-pulsar correlation described by the Hellings--Downs (HD) curve~\cite{Hellings:1983fr,Anholm:2008wy}. In 2023, the NANOGrav, EPTA/InPTA, PPTA, and CPTA collaborations independently reported evidence for a nanohertz common-spectrum process with HD-like spatial correlations~\cite{NANOGrav:2023gor,EPTA:2023fyk,Reardon:2023gzh,Xu:2023wog}. Because non-Einsteinian polarizations, modified dispersion relations, and background anisotropies all reshape this pattern~\cite{Lee2008,Lee:2010cg,Chamberlin:2011ev,Mingarelli:2013dsa,Gair:2015hra,Liang:2021bct,Liang:2023ary,AnilKumar:2023yfw,Cordes:2024oem}, PTA cross-correlations have rapidly become a powerful probe of gravitational physics in the ultralow-frequency regime~\cite{NANOGrav:2023hvm,Bernardo:2023zna,Wu:2023rib,Wang:2023div,Han:2026imx,Bernardo:2023pwt}.

From a symmetry perspective, fixed-helicity states do not exhaust the massless unitary irreducible representations of the four-dimensional Poincar\'e group. Wigner's classification also admits massless continuous-spin particles (CSPs), for which the Pauli--Lubanski invariant takes a nonzero value $W^\mu W_\mu = -\rho^2 < 0$ (in mostly-minus signature), defining a dimensionful continuous-spin scale $\rho$~\cite{Wigner:1939cj,Bekaert:2017khg}. For integer spin, a single CSP contains an infinite tower of helicity states $h\in\mathbb Z$ that mix under Lorentz boosts at finite $\rho$, analogous to the spin projections of a massive particle even though the states remain strictly lightlike, and decouple into independent fixed-helicity modes as $\rho\to 0$. Although soft theorems and the Weinberg--Witten theorem tightly constrain interacting massless particles~\cite{Weinberg:1964ew,Weinberg:1980kq} and CSPs were long regarded as non-interacting, Schuster and Toro showed that CSPs admit consistent soft-emission amplitudes and constructed a local gauge field theory on an auxiliary four-vector space $\eta^\mu$~\cite{Schuster:2013pxj,Schuster:2013vpr,Schuster:2013pta,Schuster:2014hca}, complementing related higher-spin and BRST formulations~\cite{Bekaert:2005in,Rivelles:2014fsa,Rivelles:2016rwo,Bekaert:2017xin,Najafizadeh:2017tin,Bellazzini:2024dco}. Gauge-invariant couplings of this field to worldline matter currents~\cite{Schuster:2023xqa} have recently enabled concrete predictions for continuous-spin quantum electrodynamics, atomic transitions, and photon thermodynamics~\cite{Schuster:2023jgc,Schuster:2024wjc,Reilly:2025lnm,Reilly:2025vqs}.

In the gravitational sector, a linearized theory of continuous-spin gravity with continuous-spin scale $\rho_g$ and a smooth GR limit ($\rho_g\to 0$) was recently formulated, and the time-delay signature of its $h=\pm2$ modes was evaluated for an idealized laser interferometer~\cite{Kundu:2025fsd}. In this framework, deviations from fixed-helicity dynamics enter through the dimensionless parameter $\alpha\equiv\rho_g/(\hbar\omega)$ and become more pronounced at lower frequencies $\omega$. Nanohertz gravitational waves ($f\sim 2\text{--}30\,\mathrm{nHz}$, or $\hbar\omega\sim 10^{-23}\text{--}10^{-22}\,\mathrm{eV}$) therefore probe continuous-spin scales roughly ten orders of magnitude smaller than ground-based interferometers.

At nonzero $\rho_g$, astrophysical sources may in principle radiate scalar ($h=0$), vector ($h=\pm1$), and higher-helicity ($|h|\ge3$) states alongside the tensor ($h=\pm2$) modes. Determining how a PTA responds across the entire helicity tower is thus essential both for understanding the infrared structure of continuous-spin gravity and for characterizing non-tensor polarization correlations. In this work, we derive the one-way pulsar timing redshift response and the isotropic SGWB overlap reduction function for arbitrary integer helicity $h\in\mathbb Z$ from the worldline matter coupling, correcting a trace coefficient in the massive-particle worldline current of Ref.~\cite{Kundu:2025fsd} to ensure exact recovery of GR in the $\rho_g\to 0$ limit, and extract closed-form ultraviolet and infrared scaling laws across all helicities. Focusing on an unpolarized $h=\pm2$ background, which continuously connects to GR, we then carry out a Bayesian analysis of the NANOGrav 15-year (NG15) data set, jointly sampling $\rho_g$, the common-background spectrum, and 134 intrinsic pulsar red-noise parameters.

The rest of this paper is organized as follows. In Sec.~\ref{sec:theory}, we present the worldline coupling of continuous-spin gravity and derive the one-way pulsar timing response. We then compute the overlap reduction functions for an isotropic stochastic background and analyze their asymptotic and numerical behaviors in Sec.~\ref{sec:orf}. Section~\ref{sec:data} is devoted to the Bayesian MCMC analysis of the NG15 data set and the resulting constraints on $\rho_g$. Finally, we summarize our conclusions in Sec.~\ref{sec:concl}. Throughout this paper, we adopt the metric signature $g_{\mu\nu}=\mathrm{diag}(1,-1,-1,-1)$ and use natural units with $c=\hbar=1$ unless otherwise stated.

\section{Matter coupling and pulsar timing response of continuous-spin gravity}
\label{sec:theory}
\subsection{Worldline matter coupling}

In the Schuster--Toro formulation, an integer-helicity CSP is described by a gauge field $\Psi(\eta,x)$ defined over spacetime $x^\mu$ and an auxiliary four-vector space $\eta^\mu$~\cite{Schuster:2013pta,Schuster:2014hca,Schuster:2023xqa}. Expanding the free field in plane waves of null wave four-vector $k^\mu=(\omega,\boldsymbol k)$ ($\omega=|\boldsymbol k|$), integer helicity $h\in\mathbb Z$, complex mode amplitude $a_h(\boldsymbol k)$, and helicity wave function $\psi_{h,k}(\eta)$ gives~\cite{Schuster:2023xqa,Kundu:2025fsd}
\begin{equation}\label{eq:field}
\Psi(\eta,x)=\int\frac{\dd^3\boldsymbol k}{(2\pi)^3\,2\omega}\sum_{h\in\mathbb Z}\Big[ a_h(\boldsymbol k)\psi_{h,k}(\eta)e^{-ik\cdot x}+\mathrm{c.c.}\Big].
\end{equation}
With an auxiliary null frame vector $q^\mu$ and transverse circular-polarization vectors $\epsilon_\pm^\mu$ satisfying $k\cdot q=1$, $\epsilon_+\cdot\epsilon_-=-2$, and $\epsilon_- =\epsilon_+^*$ (all other inner products vanishing), the helicity wave functions take the form~\cite{Schuster:2023xqa}
\begin{equation}\label{eq:mode}
    \psi_{h,k}(\eta)=e^{-i\rho_g\eta\cdot q}
    \begin{cases}
        (i\eta\cdot\epsilon_+)^h,&h\ge0,\\
        (-i\eta\cdot\epsilon_-)^{|h|},&h<0.
    \end{cases}
\end{equation}
Linear coupling to a conserved matter source $J(\eta,x)$ is governed by the interaction action~\cite{Schuster:2014hca,Schuster:2023xqa}
\begin{equation}\label{eq:Sint}
    S_{\rm int}=\int\dd^4x\,[\dd^4\eta]\,\delta'(\eta^2+1)J(\eta,x)\Psi(\eta,x),
\end{equation}
where $\delta'$ denotes the derivative of the Dirac $\delta$ distribution with respect to its argument and $[\dd^4\eta]$ is the regulated auxiliary measure defined in Appendix~A of Ref.~\cite{Schuster:2023xqa}.

For a point particle following worldline $\gamma^\mu(\tau)$ with tangent $u^\mu=\dd\gamma^\mu/\dd\tau$ and einbein $e(\tau)$, the matter current admits the Fourier representation~\cite{Schuster:2023xqa,Kundu:2025fsd}
\begin{equation}\label{eq:JFourier}
    J(\eta,x)=\int\frac{\dd\tau}{e(\tau)}\int\frac{\dd^4l}{(2\pi)^4}\,
    e^{-il\cdot[x-\gamma(\tau)]}j(\eta,l,u),
\end{equation}
where $j(\eta,l,u)$ is the momentum-space worldline kernel. Gauge invariance of $S_{\rm int}$ under Schuster--Toro gauge transformations requires $j(\eta,l,u)$ to satisfy the continuity condition $\delta(\eta^2+1)(-il\cdot\partial_\eta+\rho_g)j(\eta,l,u)=0$ up to terms proportional to $l\cdot u$ that integrate to total $\tau$-derivatives~\cite{Schuster:2023xqa,Kundu:2025fsd}. Recovering linearized GR as $\rho_g\to 0$ further requires $\lim_{\rho_g\to 0}j(\eta,l,u)=\frac12(\eta\cdot u)^2+\frac14 u^2$, matching the trace-reversed gravitational coupling $(\eta^\mu\eta^\nu+\frac12 g^{\mu\nu})T_{\mu\nu}$ under the $\delta'(\eta^2+1)$ measure~\cite{Schuster:2023xqa}. We note that the finite-$\rho_g$ worldline ansatz written in Sec.~III.C of Ref.~\cite{Kundu:2025fsd} carries a trace coefficient $+u^2/2$, expanding to $\frac12(\eta\cdot u)^2+\frac12 u^2$ rather than the GR kernel in their Eq.~(44). Because $(-il\cdot\partial_\eta+\rho_g)e^{-i\rho_g(\eta\cdot u)/(l\cdot u)}=0$ identically, rescaling the exponential trace term preserves worldline gauge invariance exactly. We therefore adopt the minimal gauge-invariant worldline kernel with the corrected $1/4$ trace coefficient,
\begin{equation}\label{eq:jkernel}
\begin{aligned}
j(\eta,l,u)={}&-\left(\frac{l\cdot u}{\rho_g}\right)^2 \Big[e^{-i\rho_g(\eta\cdot u)/(l\cdot u)}-1 +i\rho_g\frac{\eta\cdot u}{l\cdot u}\Big]\\
&+\frac{u^2}{4}e^{-i\rho_g(\eta\cdot u)/(l\cdot u)},
\end{aligned}
\end{equation}
whose $\rho_g\to0$ limit smoothly reproduces the GR current of Ref.~\cite{Kundu:2025fsd}. For null trajectories ($u^2=0$) and static massive observers coupled to $|h|\ge 2$ modes, the $u^2$ term drops out, leaving the interferometer tensor time delay of Ref.~\cite{Kundu:2025fsd} unaffected. In the GR limit $\rho_g\to0$, the $1/4$ trace coefficient is essential specifically to prevent an $\mathcal O(1)$ spurious scalar force and redshift response for $h=0$ (see Eq.~\eqref{eq:Q0} below), while at finite $\alpha$ it simultaneously fixes the $\mathcal O(\alpha)$ velocity coupling of massive observers to the vector modes $h=\pm1$.

We now generalize the $h=\pm2$ helicity projection of Ref.~\cite{Kundu:2025fsd} to arbitrary integer helicity. Substituting Eqs.~\eqref{eq:field} and~\eqref{eq:JFourier} into Eq.~\eqref{eq:Sint} and integrating over $x$ produces $(2\pi)^4\delta^{(4)}(k+l)$, setting $l^\mu=-k^\mu$ and giving
\begin{equation}\label{eq:Sint_mode_expanded}
\begin{aligned}
S_{\rm int}={}&\int\frac{\dd\tau}{e(\tau)}
\int\frac{\dd^3\boldsymbol k}{(2\pi)^3\,2\omega}\sum_{h\in\mathbb Z}
\int[\dd^4\eta]\,\delta'(\eta^2+1)\\
&\times\bigl[a_h(\boldsymbol k)e^{-ik\cdot\gamma(\tau)}
j(\eta,-k,u)\psi_{h,k}(\eta)+\mathrm{c.c.}\bigr].
\end{aligned}
\end{equation}
For each plane-wave mode $(h,\boldsymbol k)$, integration over $\eta$ defines the worldline helicity coupling coefficient
\begin{equation}\label{eq:Qdef}
    Q_h(u;k)\equiv
    \int[\dd^4\eta]\,
    \delta'(\eta^2+1)
    j(\eta,-k,u)\psi_{h,k}(\eta),
\end{equation}
in terms of which the single-mode interaction action reads
\begin{equation}\label{eq:Sint_single}
    S_{\mathrm{int}}^{(h,\boldsymbol k)}
    =\int\frac{\dd\tau}{e(\tau)}
    \left[
    a_h(\boldsymbol k)Q_h(u;k)e^{-ik\cdot\gamma(\tau)}
    +\mathrm{c.c.}
    \right].
\end{equation}

With the shorthand $a_h\equiv a_h(\boldsymbol k)$ at fixed wave vector, both the perturbed particle motion and the pulse redshift are determined by $Q_h(u;k)$ and its derivative $\partial Q_h/\partial u^\mu$.

For $l^\mu=-k^\mu$ and $K\equiv k\cdot u$, the on-shell worldline kernel~\eqref{eq:jkernel} reads
\begin{equation}\label{eq:j_on_shell}
\begin{aligned}
    j(\eta,-k,u)={}&-\frac{K^2}{\rho_g^2} \left( e^{i\rho_g\eta\cdot u/K} -1-\frac{i\rho_g}{K}\eta\cdot u \right)\\
    &+\frac{u^2}{4}e^{i\rho_g\eta\cdot u/K}.
\end{aligned}
\end{equation}
When Eqs.~\eqref{eq:mode} and~\eqref{eq:j_on_shell} are inserted into Eq.~\eqref{eq:Qdef}, the exponential phases $e^{i\rho_g\eta\cdot(u/K-q)}$ and $e^{-i\rho_g\eta\cdot q}$ motivate the auxiliary four-vectors
\begin{equation}\label{eq:Vdef}
    V^\mu\equiv\rho_g\left(\frac{u^\mu}{K}-q^\mu\right),
    \qquad V_0^\mu\equiv-\rho_g q^\mu.
\end{equation}

We factorize the mode function~\eqref{eq:mode} as $\psi_{h,k}(\eta)=\allowbreak e^{-i\rho_g\eta\cdot q}\chi_h(\eta)$; for $h=-n<0$, absorbing the constant sign $(-1)^n$ by redefining the complex mode amplitude $a_{-n}(\boldsymbol k)\to(-1)^n a_{-n}(\boldsymbol k)$ allows us to write $\chi_0(\eta)=1$ and $\chi_{\pm n}(\eta)=(i\eta\cdot\epsilon_\pm)^n$ uniformly for all $n\ge1$. Equation~\eqref{eq:Qdef} then splits into three integrals:
\begin{equation}\label{eq:Q_three_terms}
    \begin{aligned}
        Q_h(u;k)={}&
        \left(\frac{u^2}{4}-\frac{K^2}{\rho_g^2}\right)
        \int[\dd^4\eta]\,\delta'\,
        \chi_h(\eta)e^{i\eta\cdot V}\\
        &+\frac{K^2}{\rho_g^2}
        \int[\dd^4\eta]\,\delta'\,
        \chi_h(\eta)e^{i\eta\cdot V_0}\\
        &+\frac{iK}{\rho_g}
        \int[\dd^4\eta]\,\delta'\,(\eta\cdot u)
        \chi_h(\eta)e^{i\eta\cdot V_0},
    \end{aligned}
\end{equation}
with $\delta'\equiv\delta'(\eta^2+1)$. Each term follows in closed form from the generating integral of the Schuster--Toro formalism~\cite{Schuster:2023xqa}, which for an arbitrary four-vector $W^\mu$ reads
\begin{equation}\label{eq:generator}
    \int[\dd^4\eta]\,\delta'(\eta^2+1)
    e^{i\eta\cdot W}
    =J_0\!\left(\sqrt{-W^2}\right),
\end{equation}
where $J_n$ denotes the Bessel function of the first kind of order $n$. Applying the differential operator $(\epsilon_\pm\cdot\partial_W)^n$ ($\partial_{W^\mu}\equiv\partial/\partial W^\mu$) to Eq.~\eqref{eq:generator} brings down $(i\eta\cdot\epsilon_\pm)^n$ under the integral; using $\epsilon_\pm^2=0$ together with $J_0'(y)=-J_1(y)$ and $\frac{\dd}{\dd y}[y^{-n}J_n(y)]=-y^{-n}J_{n+1}(y)$~\cite{NISTDLMF} yields
\begin{equation}\label{eq:Bessel_derivative}
(\epsilon_\pm\cdot\partial_W)^n J_0\!\left(\sqrt{-W^2}\right) =\left(\frac{\epsilon_\pm\cdot W}{\sqrt{-W^2}}\right)^n J_n\!\left(\sqrt{-W^2}\right).
\end{equation}

In terms of the helicity projection integral
\begin{equation}\label{eq:Ghdef}
    \mathcal G_h(W)\equiv
    \int[\dd^4\eta]\,\delta'(\eta^2+1)
    \chi_h(\eta)e^{i\eta\cdot W},
\end{equation}
Eqs.~\eqref{eq:generator} and~\eqref{eq:Bessel_derivative} immediately give
\begin{align*}
\mathcal G_0(W)&=J_0\!\left(\sqrt{-W^2}\right),\\
\mathcal G_{\pm n}(W)&=\left(\frac{\epsilon_\pm\cdot W}{\sqrt{-W^2}}\right)^n J_n\!\left(\sqrt{-W^2}\right)\quad (n\ge1).
\end{align*}
The first two terms in Eq.~\eqref{eq:Q_three_terms} evaluate directly to $\mathcal G_h(V)$ and $\mathcal G_h(V_0)$, while the identity $i(\eta\cdot u)e^{i\eta\cdot W}=u^\mu\partial_{W^\mu}e^{i\eta\cdot W}$ rewrites the third integral as a directional derivative at $W=V_0$,
\[
    \int[\dd^4\eta]\,\delta'\,(\eta\cdot u)\chi_h(\eta)e^{i\eta\cdot V_0} =\frac{1}{i}u^\mu\left. \frac{\partial\mathcal G_h(W)}{\partial W^\mu}\right|_{W=V_0},
\]
yielding the coupling coefficient for arbitrary integer helicity:
\begin{equation}\label{eq:Qmaster}
\begin{aligned}
    Q_h(u;k)={}&\left(\frac{u^2}{4}-\frac{K^2}{\rho_g^2}\right) \mathcal G_h(V) +\frac{K^2}{\rho_g^2}\mathcal G_h(V_0)\\
    &+\frac{K}{\rho_g}u^\mu \left. \frac{\partial\mathcal G_h(W)}{\partial W^\mu} \right|_{W=V_0}.
\end{aligned}
\end{equation}

To express Eq.~\eqref{eq:Qmaster} explicitly in terms of Bessel functions, we introduce the dimensionless invariants
\begin{equation}\label{eq:vars}
    z\equiv\sqrt{-V^2},\qquad
    w_\pm\equiv\epsilon_\pm\cdot V
    =\frac{\rho_g(\epsilon_\pm\cdot u)}{K},
\end{equation}
where $q^2=0$ and $k\cdot q=1$ give from Eq.~\eqref{eq:Vdef}
\begin{equation}\label{eq:z2}
    z^2=\rho_g^2
    \left(\frac{2q\cdot u}{K}-\frac{u^2}{K^2}\right).
\end{equation}
Because $V_0^\mu=-\rho_g q^\mu$ satisfies $V_0^2=0$ and $\epsilon_\pm\cdot V_0=0$, the subtraction terms evaluated at $W=V_0$ depend sensitively on $|h|$. For the scalar mode ($h=0$), expanding $J_0(\sqrt{-W^2})=1+W^2/4+\mathcal O(W^4)$ about the origin gives
\[
    \mathcal G_0(V_0)=1,\qquad \left.\frac{\partial\mathcal G_0(W)}{\partial W^\mu}\right|_{W=V_0} =\frac{(V_0)_\mu}{2}=-\frac{\rho_g}{2}q_\mu,
\]
and eliminating $q\cdot u$ via Eq.~\eqref{eq:z2} yields
\begin{equation}\label{eq:Q0}
    Q_0(u;k)={} -\frac{K^2}{\rho_g^2} \left[J_0(z)-1+\frac{z^2}{4}\right] +\frac{u^2}{4}\bigl[J_0(z)-1\bigr].
\end{equation}
For the vector modes ($h=\pm1$), expanding $\mathcal G_{\pm1}(W)=(\epsilon_\pm\cdot W)J_1(\sqrt{-W^2})/\sqrt{-W^2}=\frac12(\epsilon_\pm\cdot W)+\mathcal O(W^3)$ gives $\mathcal G_{\pm1}(V_0)=0$ and
\[
    u^\mu\left.\frac{\partial\mathcal G_{\pm1}(W)}{\partial W^\mu}\right|_{W=V_0}=\frac{\epsilon_\pm\cdot u}{2}=\frac{K}{2\rho_g}w_\pm,
\]
leading to
\begin{equation}\label{eq:Q1}
    Q_{\pm1}(u;k)={} -\frac{K^2}{\rho_g^2} \left[ \frac{w_\pm}{z}J_1(z) -\frac{w_\pm}{2} \right] +\frac{u^2}{4} \frac{w_\pm}{z}J_1(z).
\end{equation}
For tensor and higher helicities ($|h|=n\ge2$), the factor $(\epsilon_\pm\cdot W)^n$ causes both $\mathcal G_{\pm n}(W)$ and its first derivative to vanish at $W=V_0$, leaving only the first term of Eq.~\eqref{eq:Qmaster}:
\begin{equation}\label{eq:Qn}
    Q_{\pm n}(u;k)=
    \left(\frac{u^2}{4}-\frac{K^2}{\rho_g^2}\right)
    \left(\frac{w_\pm}{z}\right)^n
    J_n(z).
\end{equation}

Equation~\eqref{eq:Q0} provides a direct physical check of the $u^2/4$ trace coefficient in Eq.~\eqref{eq:jkernel}. Since $J_0(z)-1=-z^2/4+\allowbreak\mathcal O(z^4)$, $Q_0(u;k)$ vanishes as $\mathcal O(\rho_g^2)$ when $\rho_g\to0$, whereas an uncorrected $u^2/2$ coefficient would leave an $\mathcal O(1)$ scalar coupling $Q_0\to u^2/4$ (and hence a nonvanishing scalar redshift response for a static observer) in the GR limit.

\subsection{One-way frequency response and timing residuals}\label{sec:oneway}

Using the single-mode coupling $Q_h(u;k)$, we vary the worldline action to obtain the linear perturbations to the photon four-momentum and the observer four-velocities, from which the one-way pulse frequency shift follows. Combining the free worldline action $S_{\mathrm{free}}=\int\dd\tau\,[u^2/(2e)+em^2/2]$ for a particle of mass $m$ ($m=0$ for photons, $m>0$ for massive observers)~\cite{Schuster:2023xqa,Kundu:2025fsd} with Eq.~\eqref{eq:Sint_single} gives $S_{\mathrm{wl}}=S_{\mathrm{free}}+S_{\mathrm{int}}^{(h,\boldsymbol k)}=\int\dd\tau\,L$, with Lagrangian
\begin{equation}\label{eq:Lworldline}
    L=
    \frac{u^2}{2e}+\frac{em^2}{2}
    +\frac{1}{e}
    \left[
    a_hQ_h(u;k)e^{-ik\cdot\gamma}
    +\mathrm{c.c.}
    \right].
\end{equation}
Variation with respect to the einbein $e(\tau)$ enforces the perturbed mass-shell constraint
\begin{equation}\label{eq:worldline_constraint}
    u^2+2\left[ a_hQ_h(u;k)e^{-ik\cdot\gamma} +\mathrm{c.c.} \right]=e^2m^2.
\end{equation}

Under worldline variations $\delta\gamma^\mu(\tau)$ to first order in $a_h$ in a constant-$e$ gauge, the canonical four-momentum
\begin{equation}\label{eq:pmom}
    p_\mu=\frac{\partial L}{\partial u^\mu} =\frac{u_\mu}{e} +\frac1e\left[ a_h\frac{\partial Q_h(u;k)}{\partial u^\mu} e^{-ik\cdot\gamma} +\mathrm{c.c.} \right]
\end{equation}
obeys the Euler--Lagrange equation $\dd p_\mu/\dd\tau=\partial L/\partial\gamma^\mu$,
\begin{equation}\label{eq:EL}
    \frac{\dd p_\mu}{\dd\tau}
    =-\frac{ia_h}{e}k_\mu Q_h(u;k)e^{-ik\cdot\gamma}
    +\mathrm{c.c.}
\end{equation}
Evaluating the right-hand side along the unperturbed straight worldline $\gamma_0^\mu(\tau)$ (where $u^\mu$ and $K=k\cdot u$ are constant) and integrating via $\dd e^{-ik\cdot\gamma_0}/\dd\tau=-iKe^{-ik\cdot\gamma_0}$ gives the canonical-momentum perturbation
\begin{equation}\label{eq:dp}
    \delta p_\mu=\frac{a_h}{e}
    \frac{k_\mu}{k\cdot u}Q_h(u;k)e^{-ik\cdot\gamma_0}
    +\mathrm{c.c.},
\end{equation}
with integration constants canceling between reception and emission in a one-way timing link for an intrinsically steady clock. Comparison of Eq.~\eqref{eq:dp} with the linearization of Eq.~\eqref{eq:pmom} isolates the tangent-vector perturbation:
\begin{equation}\label{eq:du}
    \delta u_\mu=a_h\left[\frac{k_\mu}{k\cdot u}Q_h(u;k) -\frac{\partial Q_h(u;k)}{\partial u^\mu}\right] e^{-ik\cdot\gamma_0}+\mathrm{c.c.},
\end{equation}
comprising a term along the wave four-vector $k_\mu$ and a derivative term $-\partial Q_h/\partial u^\mu$ arising from the velocity dependence of the coupling.

Because the photon frequency $\nu$ measured by an observer with four-velocity $U^\mu$ ($U^2=1$) equals the invariant contraction $p_{\gamma\mu}U^\mu$~\cite{Wald:1984rg,Kundu:2025fsd}, linearizing about an unperturbed null ray of four-momentum $p_{\gamma\mu}^{(0)}=\nu_0 v_\mu$ ($v^2=0$, $v\cdot U=1$, $\nu_0=p_{\gamma\mu}^{(0)}U^\mu$, constant einbein $e_\gamma=\nu_0^{-1}$) and observer velocity $U^\mu+\delta U^\mu$ yields the fractional shift
\begin{equation}\label{eq:frequency_linear}
    \frac{\delta\nu}{\nu}
    =\frac{\delta p_{\gamma\mu}U^\mu
        +p_{\gamma\mu}^{(0)}\delta U^\mu}
    {p_{\gamma\mu}^{(0)}U^\mu}.
\end{equation}
At an emission or reception event $x^\mu$, evaluating Eq.~\eqref{eq:dp} on the photon ($u=v$) gives the photon-momentum contribution
\begin{equation}\label{eq:frequency_photon_part}
    \frac{\delta p_{\gamma\mu}U^\mu}{\nu_0}
    =a_h e^{-ik\cdot x}\frac{k\cdot U}{k\cdot v}Q_h(v;k)
    +\mathrm{c.c.},
\end{equation}
which is invariant under affine reparameterization by the quadratic homogeneity of $Q_h(v;k)$, while evaluating Eq.~\eqref{eq:du} on the massive observer ($u=U$, $e_U=m^{-1}$) gives the observer-velocity contribution
\begin{equation}\label{eq:frequency_observer_part}
\begin{aligned}
\frac{p_{\gamma\mu}^{(0)}\delta U^\mu}{\nu_0}={}&v_\mu\delta U^\mu\\
={}&a_h e^{-ik\cdot x}\biggl[ \frac{k\cdot v}{k\cdot U}Q_h(U;k)\\
&\qquad\qquad -v^\mu\frac{\partial Q_h(U;k)}{\partial U^\mu}\biggr]+\mathrm{c.c.}
\end{aligned}
\end{equation}
The sum of Eqs.~\eqref{eq:frequency_photon_part} and~\eqref{eq:frequency_observer_part} expresses the single-mode local frequency shift at $x^\mu$ as
\begin{equation}\label{eq:frequency_one_mode}
    \frac{\delta\nu_h}{\nu}
    =a_h e^{-ik\cdot x}\,\mathcal D_h(v,U)
    +\mathrm{c.c.},
\end{equation}
with directional response coefficient
\begin{equation}\label{eq:D}
\begin{aligned}
    \mathcal D_h(v,U)={}&\frac{k\cdot U}{k\cdot v}Q_h(v;k) +\frac{k\cdot v}{k\cdot U}Q_h(U;k)\\
    &-v^\mu\frac{\partial Q_h(U;k)}{\partial U^\mu}.
\end{aligned}
\end{equation}

To evaluate Eq.~\eqref{eq:D} explicitly in the observer rest frame $U^\mu=(1,0,0,0)$, we align the gravitational wave along $+\hat{\boldsymbol z}$:
\[
\begin{gathered}
    k^\mu=\omega(1,0,0,1),\qquad
    q^\mu=\frac{1}{2\omega}(1,0,0,-1),\\
    \epsilon_\pm^\mu=(0,1,\pm i,0),
    \qquad U^\mu=(1,0,0,0).
\end{gathered}
\]
Parameterizing the photon four-vector in spherical angles $(\theta,\varphi)$ as $v^\mu=(1,\allowbreak\sin\theta\cos\varphi,\allowbreak\sin\theta\sin\varphi,\allowbreak\cos\theta)$ with $\mu\equiv\cos\theta$ and $\alpha\equiv\rho_g/\omega$, we have $k\cdot v=\omega(1-\mu)$, $q\cdot v=(1+\mu)/(2\omega)$, and $v^2=0$; Eqs.~\eqref{eq:vars} and~\eqref{eq:z2} at $u=v$ then give
\begin{equation}\label{eq:zsky}
    z=\alpha\sqrt{\frac{1+\mu}{1-\mu}},
    \qquad
    \frac{w_\pm}{z}=-e^{\pm i\varphi}.
\end{equation}
The dimensionless Bessel argument $z=\alpha\cot(\theta/2)$ in Eq.~\eqref{eq:zsky} admits a direct physical interpretation in terms of Wigner's $\mathrm{ISO}(2)$ little-group kinematics. Since $z=|w_\pm|=\rho_g|\epsilon_\pm\cdot v|/(k\cdot v)$ for $v^2=0$, it measures the ratio of the continuous-spin scale projected transverse to the photon trajectory ($\propto\sin\theta$) to the rate at which the gravitational-wave phase varies along the photon path ($\propto 1-\cos\theta$). For head-on propagation ($\theta=\pi$, $\mu=-1$), the vanishing transverse projection eliminates little-group mixing ($z=0$). Conversely, for nearly co-propagating waves and photons ($\theta\to0$, $\mu\to1^-$, corresponding to pulsars near $-\hat{\boldsymbol k}$ with $\hat{\boldsymbol k}\cdot\hat{\boldsymbol p}\to-1$), prolonged phase coherence causes the helicity-mixing argument $z$ to grow large within a narrow collinear region $1-\mu\lesssim\alpha^2$, where $z\gtrsim1$ and the helicity tower mixes strongly even when $\alpha\ll1$.

For the static-observer terms in Eq.~\eqref{eq:D}, setting $u=U=(1,0,0,0)$ gives $\epsilon_\pm\cdot U=0$ and hence $w_\pm(U)=0$, while $k\cdot U=\omega$, $q\cdot U=1/(2\omega)$, and $U^2=1$ imply $z(U)=0$, so that $Q_h(U;k)=0$ identically for every $h\in\mathbb Z$. Differentiating Eqs.~\eqref{eq:Q0}--\eqref{eq:Qn} with respect to the contravariant tangent $u^\mu$ prior to setting $u=U$, and using $\partial w_\pm/\partial u^\mu|_{u=U}=(\rho_g/\omega)\epsilon_{\pm\mu}$ together with the small-argument Bessel expansions, yields
\begin{equation}\label{eq:Ueval}
    \left.\frac{\partial Q_h(u;k)}{\partial u^\mu}\right|_{u=U}
    =\begin{cases}
        \displaystyle\frac{\rho_g}{8\omega}
        \epsilon_{\pm\mu},&h=\pm1,\\[2mm]
        0,&h=0\text{ or }|h|\ge2.
    \end{cases}
\end{equation}
Equation~\eqref{eq:Ueval} reflects an angular-momentum selection rule in the observer rest frame. Because $U^\mu=(1,\boldsymbol 0)$ is rotationally invariant about $\hat{\boldsymbol k}$, a linear spatial velocity perturbation $\delta\boldsymbol U_\perp$ carrying azimuthal spin weight $\pm1$ cannot be excited by $|h|\ge2$ modes, while the gauge subtractions in Eq.~\eqref{eq:Q0} cancel any longitudinal velocity perturbation for $h=0$, leaving only the vector modes $h=\pm1$ to induce a transverse velocity perturbation of a static observer. Contracting Eq.~\eqref{eq:Ueval} with $-v^\mu$ and using $v^\mu\epsilon_{\pm\mu}=-\sqrt{1-\mu^2}\,e^{\pm i\varphi}$ gives the nonvanishing vector observer-velocity contribution,
\begin{equation}\label{eq:D1endpoint}
    -v^\mu\left.
    \frac{\partial Q_{\pm1}(u;k)}
    {\partial u^\mu}
    \right|_{u=U}
    =\frac{\alpha}{8}
    \sqrt{1-\mu^2}\,e^{\pm i\varphi}.
\end{equation}

Substitution of $k\cdot U=\omega$, $k\cdot v=\omega(1-\mu)$, and $Q_h(U;k)=0$ into Eq.~\eqref{eq:D} reduces the directional response to
\begin{equation}\label{eq:Dsky}
    \mathcal D_h(v,U)
    =\frac{Q_h(v;k)}{1-\mu}
    -v^\mu\left.
    \frac{\partial Q_h(u;k)}{\partial u^\mu}
    \right|_{u=U}.
\end{equation}
Evaluating $Q_h(v;k)$ at $u=v$ ($v^2=0$) via Eqs.~\eqref{eq:Q0}--\eqref{eq:Qn} together with Eqs.~\eqref{eq:Ueval}--\eqref{eq:D1endpoint} yields the local frequency-shift coefficients across the entire helicity tower:
\begin{align}
    \mathcal D_0
    &=-\frac{1-\mu}{\alpha^2}
    \left[J_0(z)-1+\frac{z^2}{4}\right],
    \label{eq:D0}\\
    \mathcal D_{\pm1}
    &=e^{\pm i\varphi}
    \left[
    \frac{1-\mu}{\alpha^2}
    \left(J_1(z)-\frac z2\right)
    +\frac{\alpha}{8}\sqrt{1-\mu^2}
    \right],
    \label{eq:D1}\\
    \mathcal D_{\pm n}
    &=(-1)^{n+1}e^{\pm in\varphi}
    \frac{1-\mu}{\alpha^2}J_n(z),
    \qquad n\ge2.
    \label{eq:Dn}
\end{align}
Every response carries the azimuthal phase $e^{ih\varphi}$; only $h=\pm1$ receives contributions from both the photon-momentum and observer-velocity perturbations, whereas all $h=0$ and $|h|\ge2$ responses arise solely from the photon-momentum perturbation.

Following standard PTA antenna-pattern conventions~\cite{Anholm:2008wy,Romano:2023zhb}, we factorize the antenna response $F_h$ into its azimuthal phase and polar profile $A_{|h|}(\mu;\alpha)$ via
\begin{equation}\label{eq:Fgeneral}
    F_h(\mu,\varphi;\alpha)\equiv-4(-1)^{|h|}\mathcal D_h(v,U) =e^{ih\varphi}A_{|h|}(\mu;\alpha),
\end{equation}
giving
\begin{align}
A_0(\mu;\alpha)&=\frac{4(1-\mu)}{\alpha^2}
\left[J_0(z)-1+\frac{z^2}{4}\right],\label{eq:A0}\\
A_1(\mu;\alpha)&=\frac{4(1-\mu)}{\alpha^2}\left[J_1(z)-\frac z2\right]
+\frac{\alpha}{2}\sqrt{1-\mu^2},\label{eq:A1}\\
A_n(\mu;\alpha)&=\frac{4(1-\mu)}{\alpha^2}J_n(z),\qquad n\ge2.\label{eq:An}
\end{align}
Despite the divergence of $z$ as $\mu\to1^-$, every $A_{|h|}(\mu;\alpha)$ is continuous and bounded across $\mu\in[-1,1]$. At head-on incidence ($\mu=-1$, $z=0$), $A_{|h|}(-1;\alpha)=0$ for all $h$, while in the collinear limit ($\mu\to1^-$, $z\to\infty$ at fixed $\alpha>0$), $A_0\to2$ whereas $A_{n\ge1}\to0$. Conversely, at any fixed $\mu<1$, taking $\alpha\to0$ ($z\to0$) gives $A_2(\mu;\alpha)\to\frac12(1+\mu)$, which reproduces the standard GR circular-polarization antenna patterns $F_{\pm2}$, while all non-tensor profiles $A_{n\neq2}(\mu;\alpha)$ vanish pointwise on $\mu\in[-1,1)$ as well as in the $L^2(S^2)$ norm (the narrow collinear region $1-\mu\lesssim\alpha^2$ subtending a vanishing solid angle $\Delta\Omega\sim\alpha^2$).

For pulsar $a$ at distance $L_a$ along unit sky vector $\hat{\boldsymbol p}_a$ from Earth (placed at the spatial origin), radio pulses propagate along $\hat{\boldsymbol v}_a=-\hat{\boldsymbol p}_a$ with $v_a^\mu=(1,\hat{\boldsymbol v}_a)$ and directional cosine $\mu_a=\hat{\boldsymbol k}\cdot\hat{\boldsymbol v}_a=-\hat{\boldsymbol k}\cdot\hat{\boldsymbol p}_a$ relative to $\hat{\boldsymbol k}=\boldsymbol k/\omega$. Relating the pulsar emission event $x_{\mathrm P}^\mu=(t-L_a,L_a\hat{\boldsymbol p}_a)$ to the Earth reception event $x_{\mathrm E}^\mu=(t,\boldsymbol0)$ for a stable intrinsic spin frequency gives the one-way fractional frequency shift
\begin{equation}\label{eq:link_frequency_difference}
    \frac{\delta\nu_{a,h}(t)}{\nu_0} ={}a_h\mathcal D_h(v_a,U) \left( e^{-ik\cdot x_{\mathrm E}} -e^{-ik\cdot x_{\mathrm P}} \right) +\mathrm{c.c.},
\end{equation}
where $k\cdot x_{\mathrm E}=\omega t$ and $k\cdot x_{\mathrm P}=\omega t-\Phi_a$, with pulsar phase $\Phi_a\equiv\omega L_a(1+\hat{\boldsymbol k}\cdot\hat{\boldsymbol p}_a)=\omega L_a(1-\mu_a)$, yielding the Earth- and pulsar-term phase factor $e^{-i\omega t}(1-e^{i\Phi_a})$~\cite{Estabrook:1975jtn,Anholm:2008wy,Romano:2023zhb}.

In terms of the pulse redshift $z_a(t)\equiv-\delta\nu_a(t)/\nu_0$, the timing residual $r_a(t)$ accumulates as~\cite{Detweiler:1979wn,Anholm:2008wy}
\begin{equation}\label{eq:timing_residual_integral}
    r_a(t)=\int^t z_a(t')\,\dd t'.
\end{equation}
We adopt the two-sided Fourier convention $\widetilde z_a(f)=\int_{-\infty}^{\infty}\dd t\,e^{i2\pi ft}z_a(t)$ ($\omega=2\pi f$), where the dimensionless parameter entering the antenna patterns across all frequencies is defined as the positive quantity $\alpha\equiv\rho_g/|\omega|$ (reducing to $\rho_g/\omega$ for $f>0$). This definition ensures that the odd Bessel and observer-velocity contributions in $A_{2k+1}$ do not flip signs at negative frequencies, guaranteeing $F_h^*(\hat{\boldsymbol k};\alpha)=F_{-h}(\hat{\boldsymbol k};\alpha)$. Absorbing the overall redshift sign, the plane-wave integration measure, and the prefactor of Eq.~\eqref{eq:Fgeneral} into the plane-wave amplitude $b_h(f,\hat{\boldsymbol k})$, the reality of $z_a(t)$ then implies the standard circular-polarization reality condition $b_{-h}(f,\hat{\boldsymbol k})=b_h^*(-f,\hat{\boldsymbol k})$. Summing over propagation directions and helicities then gives the frequency-domain redshift
\begin{equation}\label{eq:full_link_redshift}
    \widetilde z_a(f) ={}\sum_{h\in\mathbb Z}\int_{S^2}\dd\Omega_{\hat{\boldsymbol k}}\, b_h(f,\hat{\boldsymbol k}) F_{a,h}(\hat{\boldsymbol k};\alpha) \left(1-e^{i\Phi_a}\right),
\end{equation}
where $F_{a,h}(\hat{\boldsymbol k};\alpha)\equiv F_h(\mu_a,\varphi_a;\alpha)$ and $\varphi_a$ is the azimuthal angle of $\hat{\boldsymbol v}_a$ around $\hat{\boldsymbol k}$. With $\widetilde r_a(f)=\widetilde z_a(f)/\allowbreak(-i\omega)$, in the short-wavelength regime $|\omega|L_a\gg1$~\cite{Anholm:2008wy,Mingarelli:2018kgp} the rapidly oscillating pulsar phase $e^{i\Phi_a}$ averages out of distinct-pulsar cross-correlations ($a\neq b$) while doubling each zero-lag autocorrelation ($a=b$). Retaining the coherent Earth-term timing residual gives
\begin{equation}\label{eq:residualEarth}
    \widetilde r_{a,\mathrm E}(f) ={}\frac{1}{-i\omega} \sum_{h\in\mathbb Z} \int_{S^2}\dd\Omega_{\hat{\boldsymbol k}}\, b_h(f,\hat{\boldsymbol k}) F_{a,h}(\hat{\boldsymbol k};\alpha).
\end{equation}

\section{Angular correlations of a stochastic continuous-spin background}
\label{sec:orf}

\subsection{Multipole expansion of the overlap reduction function}
\label{sec:orf_multipole}

Consider a zero-mean, stationary, homogeneous, and isotropic stochastic continuous-spin background~\cite{Allen:1997ad,Anholm:2008wy,Romano:2016dpx}. Because a rotation by $\chi$ about $\hat{\boldsymbol k}$ transforms $b_h(f,\hat{\boldsymbol k})\to e^{ih\chi}b_h(f,\hat{\boldsymbol k})$, statistical isotropy requires a rotationally invariant two-point function that is diagonal in helicity:
\begin{equation}\label{eq:cov_diagonal}
    \av{b_h(f,\hat{\boldsymbol k}) b_{h'}^*(f',\hat{\boldsymbol k}')} =\frac{S_h(f)}{4\pi} \delta_{hh'}\delta(f-f') \delta^{(2)}(\hat{\boldsymbol k},\hat{\boldsymbol k}'),
\end{equation}
where angular brackets denote ensemble expectation, $\delta_{hh'}$ is the Kronecker delta, $\delta^{(2)}(\hat{\boldsymbol k},\hat{\boldsymbol k}')$ is the Dirac distribution on $S^2$, and $S_h(f)\ge0$ is the polarization power spectral density of helicity $h$.

Substituting Eq.~\eqref{eq:residualEarth} into the cross-correlation of two distinct pulsars ($a\ne b$) separated by angle $\zeta=\arccos(\hat{\boldsymbol p}_a\cdot\hat{\boldsymbol p}_b)=\arccos(\hat{\boldsymbol v}_a\cdot\hat{\boldsymbol v}_b)$ and using Eq.~\eqref{eq:cov_diagonal}, we define the Earth-term overlap reduction function $\Gamma_h^{\mathrm E}(\zeta;\alpha)$ ($\alpha=\rho_g/\omega$) as
\begin{equation}\label{eq:GammaEarth}
    \Gamma_h^{\mathrm E}(\zeta;\alpha)
    \equiv\frac34
    \int_{S^2}\frac{\dd\Omega_{\hat{\boldsymbol k}}}{4\pi}\,
    F_{a,h}(\hat{\boldsymbol k};\alpha)
    F_{b,h}^*(\hat{\boldsymbol k};\alpha),
\end{equation}
normalized by $3/4$ so that summing the two tensor helicities ($h=\pm2$) recovers the standard HD normalization at $\alpha\to0$. The Earth-term timing-residual cross-correlation then factorizes into source spectrum and detector geometry:
\begin{equation}\label{eq:cross_final}
    \av{\widetilde r_{a,\mathrm E}(f)\widetilde r_{b,\mathrm E}^*(f')} =\frac{4\,\delta(f-f')}{3\omega^2}\sum_{h\in\mathbb Z} S_h(f)\Gamma_h^{\mathrm E}(\zeta;\alpha).
\end{equation}

Exploiting the azimuthal factorization $F_h(\mu,\varphi;\alpha)=e^{ih\varphi}A_{|h|}(\mu;\alpha)$, we convert the two-dimensional sky integral~\eqref{eq:GammaEarth} into an exact Legendre series via spherical-harmonic projection and $\mathrm{SO}(3)$ orthogonality~\cite{Varshalovich:1988ifq,Mingarelli:2013dsa,Gair:2014rwa,Gair:2015hra,Bernardo:2022rif,Allen:2024bnk}. In the reference frame $\hat{\boldsymbol k}=\hat{\boldsymbol z}$, expanding $F_h(\hat{\boldsymbol v};\alpha)=\allowbreak\sum_{\ell=0}^{\infty}\allowbreak\sum_{m=-\ell}^{\ell}\allowbreak a_{\ell m}^{\mathrm E,(h)}(\alpha)\allowbreak Y_{\ell m}(\hat{\boldsymbol v})$ in the Condon--Shortley spherical harmonics ($m\ge0$),
\[
    Y_{\ell m}(\theta,\varphi)
    =\sqrt{\frac{2\ell+1}{4\pi}\frac{(\ell-m)!}{(\ell+m)!}}\,
    P_\ell^m(\cos\theta)e^{im\varphi},
\]
where $P_\ell^m(\mu)$ is the associated Legendre function~\cite{NISTDLMF}, motivates defining the one-dimensional multipole projection integral for $n=|h|$ and $\ell\ge n$:
\begin{equation}\label{eq:IlE}
    I_{\ell,\mathrm E}^{(n)}(\alpha)
    \equiv\sqrt{\frac{(\ell-n)!}{(\ell+n)!}}
    \int_{-1}^{1}\dd\mu\,
    P_\ell^n(\mu)A_n(\mu;\alpha).
\end{equation}
Azimuthal integration yields $2\pi\delta_{mh}$, forcing $a_{\ell m}^{\mathrm E,(h)}(\alpha)=0$ for $\ell<|h|$ and giving, for $h\ge0$ and $\ell\ge h$,
\begin{equation}\label{eq:harmonic_response_coefficient}
\begin{aligned}
    a_{\ell m}^{\mathrm E,(h)}(\alpha) &=\int_{S^2}\dd\Omega_{\hat{\boldsymbol v}}\, Y_{\ell m}^*(\hat{\boldsymbol v})F_h(\hat{\boldsymbol v};\alpha)\\
    &=\delta_{mh}\sqrt{\pi(2\ell+1)}\, I_{\ell,\mathrm E}^{(h)}(\alpha).
\end{aligned}
\end{equation}
If we write $a_{\ell h}^{\mathrm E}(\alpha)\equiv a_{\ell h}^{\mathrm E,(h)}(\alpha)$ for the single nonvanishing azimuthal component, the conjugation identity $Y_{\ell,-n}=(-1)^n Y_{\ell n}^*$ for $h=-n<0$ contributes at most an overall sign $(-1)^n$, so that $|a_{\ell,\pm n}^{\mathrm E}(\alpha)|^2=\pi(2\ell+1)|I_{\ell,\mathrm E}^{(n)}(\alpha)|^2$.

A rotation $R\in\mathrm{SO}(3)$ mapping $\hat{\boldsymbol z}$ to a general wave direction $\hat{\boldsymbol k}=R\hat{\boldsymbol z}$ transforms $Y_{\ell h}(R^{-1}\hat{\boldsymbol v})=\allowbreak\sum_{m=-\ell}^{\ell}\allowbreak D_{mh}^{\ell}(R)Y_{\ell m}(\hat{\boldsymbol v})$ via Wigner $D$ matrices~\cite{Varshalovich:1988ifq}. Because any residual rotation about $\hat{\boldsymbol k}$ produces opposite helicity phases that cancel in $F_{a,h}F_{b,h}^*$, averaging $\hat{\boldsymbol k}$ over $S^2$ in Eq.~\eqref{eq:GammaEarth} is equivalent to integrating over $\mathrm{SO}(3)$ with normalized Haar measure $\dd\bar R$ ($\int_{\mathrm{SO}(3)}\dd\bar R=1$). Combining the Wigner orthogonality relation~\cite{Varshalovich:1988ifq}
\begin{equation}\label{eq:wigner_rotation_orthogonality}
    \int_{\mathrm{SO}(3)}\dd\bar R\,
    D_{mh}^{\ell}(R)D_{m'h}^{\ell' *}(R)
    =\frac{\delta_{\ell\ell'}\delta_{mm'}}{2\ell+1}
\end{equation}
with the spherical-harmonic addition theorem $\sum_{m=-\ell}^{\ell}\allowbreak Y_{\ell m}(\hat{\boldsymbol v}_a)\allowbreak Y_{\ell m}^*(\hat{\boldsymbol v}_b)=\allowbreak\frac{2\ell+1}{4\pi}\allowbreak P_\ell(\hat{\boldsymbol v}_a\cdot\hat{\boldsymbol v}_b)$, where $P_\ell$ denotes the Legendre polynomial of degree $\ell$, reduces the sky average to
\begin{equation}\label{eq:sky_response_rotation_average}
    \int_{S^2}\frac{\dd\Omega_{\hat{\boldsymbol k}}}{4\pi} F_{a,h}(\hat{\boldsymbol k};\alpha)F_{b,h}^*(\hat{\boldsymbol k};\alpha) =\sum_{\ell=|h|}^{\infty}\frac{|a_{\ell h}^{\mathrm E}(\alpha)|^2}{4\pi} P_\ell(\cos\zeta).
\end{equation}
Substitution of $|a_{\ell h}^{\mathrm E}(\alpha)|^2$ together with the normalization factor $3/4$ from Eq.~\eqref{eq:GammaEarth} yields the Earth-term angular correlation series for arbitrary integer helicity:
\begin{equation}\label{eq:masterE}
\Gamma_h^{\mathrm E}(\zeta;\alpha)
=\frac{3}{16}\sum_{\ell=|h|}^{\infty}(2\ell+1)
\left|I_{\ell,\mathrm E}^{(|h|)}(\alpha)\right|^2P_\ell(\cos\zeta).
\end{equation}

Several structural properties follow directly from Eq.~\eqref{eq:masterE}. First, azimuthal spin weight imposes the multipole selection rule $\ell\ge|h|$, forbidding a monopole ($\ell=0$) for all $h\neq0$ and both monopole and dipole ($\ell=0,1$) for $|h|\ge2$. Second, since $I_{\ell,\mathrm E}^{(|h|)}(\alpha)$ depends solely on $|h|$, opposite helicities share identical spatial correlations, $\Gamma_{+n}^{\mathrm E}(\zeta;\alpha)=\Gamma_{-n}^{\mathrm E}(\zeta;\alpha)$. Third, the manifest non-negativity of the Legendre weights $(2\ell+1)|I_{\ell,\mathrm E}^{(|h|)}(\alpha)|^2\ge0$ guarantees, by Schoenberg's theorem~\cite{Schoenberg1942}, that the inter-pulsar correlation matrix on $S^2$ is positive semi-definite for every helicity $h$ and every scale ratio $\alpha$. Finally, as $\alpha\to0$, $A_2(\mu;\alpha)\to\frac12(1+\mu)$ gives $|I_{\ell,\mathrm E}^{(2)}(0)|^2=4/[(\ell-1)\ell(\ell+1)(\ell+2)]$ for $\ell\ge2$, so the unpolarized tensor correlation $\Gamma_T^{\mathrm E}(\zeta;\alpha)\equiv\Gamma_{+2}^{\mathrm E}(\zeta;\alpha)+\Gamma_{-2}^{\mathrm E}(\zeta;\alpha)=2\Gamma_{+2}^{\mathrm E}(\zeta;\alpha)$ resums to the GR Hellings--Downs curve~\cite{Hellings:1983fr,Romano:2023zhb,Allen:2024bnk},
\begin{equation}\label{eq:HD}
\Gamma_{\rm HD}(\zeta)=\frac12+\frac32\xi\ln\xi-\frac14\xi,\qquad \xi\equiv\frac{1-\cos\zeta}{2},
\end{equation}
with $\Gamma_{\rm HD}(0)=1/2$ for distinct pulsars and unit diagonal entries once the pulsar term is added.

\subsection{Analytical ultraviolet and infrared scaling laws}
\label{sec:asymptotics}

\begin{table*}[t]
\caption{Summary of multipole selection rules, static-observer velocity perturbation $\delta U^\mu$, and Earth-term ultraviolet and infrared asymptotic behaviors ($\Gamma_0^{\mathrm E}$ for $h=0$, unpolarized sum $2\Gamma_{+n}^{\mathrm E}$ for $|h|=n\ge1$; entries without explicit $\zeta$ refer to zero lag $\zeta=0$) across the continuous-spin helicity tower.}
\label{tab:scaling_summary}
\centering
\footnotesize
\setlength{\tabcolsep}{6pt}
\renewcommand{\arraystretch}{1.3}
\begin{tabular*}{\textwidth}{@{\extracolsep{\fill}}lllll@{}}
\toprule
Sector & $\ell$ range & Observer $\delta U^\mu$ & UV ($\alpha\ll1$) & IR behavior ($\alpha\gg 1$) \\
\midrule
Scalar ($h=0$) & $\ell\ge0$ & $0$ & $\frac{19}{72}\alpha^2$ & $\frac34+\frac14\cos\zeta$ \\
Vector ($h=\pm1$) & $\ell\ge1$ & $\propto\alpha\epsilon_\pm^\mu$ & $\frac34\alpha^2\ln(1/\alpha)$ & $\frac14(\alpha-4/\alpha)^2\cos\zeta$ \\
Tensor ($h=\pm2$) & $\ell\ge2$ & $0$ & $\frac12-\frac12\alpha^2\ln(1/\alpha)$ & $30/\alpha^5$ ($\zeta=0$); $\ell\sim\mathcal O(\alpha)$ \\
Higher ($h=\pm3$) & $\ell\ge3$ & $0$ & $\frac{1}{12}\alpha^2\ln(1/\alpha)$ & $30/\alpha^5$ ($\zeta=0$); $\ell\sim\mathcal O(\alpha)$ \\
Higher ($|h|=n\ge4$) & $\ell\ge n$ & $0$ & $\frac{30\,(n-4)!}{(n+3)!}\alpha^2$ & $30/\alpha^5$ ($\zeta=0$); $\ell\sim\mathcal O(\alpha)$ \\
\bottomrule
\end{tabular*}
\vspace{1ex}
\raggedright\footnotesize At intermediate coupling $\alpha\sim\mathcal O(n)$, the zero-lag correlation $2\Gamma_{+n}^{\mathrm E}(0;\alpha)$ for $|h|\ge3$ exhibits a non-monotonic peak (e.g., $\alpha\simeq{1.2}$ for $|h|=3$) before entering the universal $30/\alpha^5$ asymptotic decay. For $|h|=n\ge2$, the infrared asymptotics require $\alpha\gg n$.
\end{table*}

Before turning to numerical evaluations, we extract closed-form asymptotic scaling laws from Eqs.~\eqref{eq:A0}--\eqref{eq:An} and~\eqref{eq:masterE} that govern the high-frequency ultraviolet ($\alpha\ll1$) and deep infrared ($\alpha\gg1$) limits across the entire helicity tower. Under the change of variables $x\equiv\sqrt{(1+\mu)/(1-\mu)}=\cot(\theta/2)\in[0,\infty)$, for which $z=\alpha x$, $1-\mu=2/(x^2+1)$, and $\dd\mu=4x\,\dd x/(x^2+1)^2$, Parseval's identity in Eq.~\eqref{eq:masterE} expresses the zero-separation Earth-term correlations as exact one-dimensional integrals; in particular, for $|h|=n\ge2$,
\begin{equation}\label{eq:zero_lag_xintegral}
\begin{aligned}
    2\Gamma_{+n}^{\mathrm E}(0;\alpha)
    &=\frac34\int_{-1}^{1}\dd\mu\,A_n(\mu;\alpha)^2\\
    &=\frac{192}{\alpha^4}\int_0^{\infty}\frac{x\,J_n(\alpha x)^2}{(x^2+1)^4}\,\dd x.
\end{aligned}
\end{equation}

In the ultraviolet regime ($\alpha\ll1$), naive pointwise expansion of $J_n(\alpha x)\propto(\alpha x)^n$ at fixed $\mu<1$ would suggest $A_n\propto\alpha^{n-2}$ and hence sky-averaged power scaling as $\alpha^{2n-4}$ for $n\ge3$. For $n\ge3$, however, the resulting $x$-integrand behaves as $x^{2n+1}/(x^2+1)^4\sim x^{2n-7}$ at large $x$, diverging toward the collinear region $x\gtrsim\alpha^{-1}$ ($1-\mu\lesssim\alpha^2$) where $z=\alpha x\sim\mathcal O(1)$ and Bessel oscillations regulate the integral. Consequently, this collinear region dominates the sky average, lifting all $|h|\ge4$ correlations to $\mathcal O(\alpha^2)$ and generating logarithmic $\alpha^2\ln(1/\alpha)$ terms for $|h|=1,2,3$. Asymptotic expansion of Eq.~\eqref{eq:zero_lag_xintegral} and the corresponding $h=0,\pm1$ integrals for $\alpha\ll1$ yields
\begin{align}
    \Gamma_0^{\mathrm E}(0;\alpha)
    &=\frac{19}{72}\alpha^2+\mathcal O(\alpha^4\ln\alpha^{-1}),\label{eq:UV_h0}\\
    2\Gamma_{+1}^{\mathrm E}(0;\alpha)
    &=\frac34\alpha^2\ln\frac{1}{\alpha}+\mathcal O(\alpha^2),\label{eq:UV_h1}\\
    2\Gamma_{+2}^{\mathrm E}(0;\alpha)
    &=\frac12-\frac{\alpha^2}{2}\left(\ln\frac{2}{\alpha}-\gamma_{\mathrm E}+\frac{13}{120}\right)\nonumber\\
    &\quad +\mathcal O(\alpha^4\ln\alpha^{-1}),\label{eq:UV_h2}\\
    2\Gamma_{+3}^{\mathrm E}(0;\alpha)
    &=\frac{1}{12}\alpha^2\ln\frac{1}{\alpha}+\mathcal O(\alpha^2),\label{eq:UV_h3}\\
    2\Gamma_{+n}^{\mathrm E}(0;\alpha)
    &=\frac{30\,(n-4)!}{(n+3)!}\,\alpha^2+\mathcal O(\alpha^4\ln\alpha^{-1}),
    \quad n\ge4,\label{eq:UV_hn}
\end{align}

where $\gamma_{\mathrm E}\simeq0.5772$ is the Euler--Mascheroni constant, so that the full tensor autocorrelation $D_T(\alpha)=4\Gamma_{+2}^{\mathrm E}(0;\alpha)$ satisfies $D_T(\alpha)\simeq 1-\alpha^2[\ln(1/\alpha)+{0.2243}]$. Averaging over an isotropic pulsar sky thus enhances the leading continuous-spin correction to the tensor sector by $\ln(1/\alpha)$ relative to fixed-angle interferometer links~\cite{Kundu:2025fsd}.

In the deep infrared regime ($\alpha\gg1$), the helicity modes exhibit three distinct asymptotic behaviors depending on $|h|$. For the scalar mode ($h=0$), the bounded oscillatory piece $4(1-\mu)[J_0(z)-1]/\alpha^2$ in Eq.~\eqref{eq:A0} decays as $\mathcal O(\alpha^{-2})$ away from $\mu=1$, leaving $A_0(\mu;\alpha)\to 1+\mu$ so that the sky correlation approaches a positive monopole-plus-dipole limit,
\begin{equation}\label{eq:IR_h0}
    \lim_{\alpha\to\infty}\Gamma_0^{\mathrm E}(\zeta;\alpha)
    =\frac34+\frac14\cos\zeta.
\end{equation}
For the vector modes ($h=\pm1$), rewriting Eq.~\eqref{eq:A1} as $A_1(\mu;\alpha)=4(1-\mu)J_1(z)/\alpha^2+(\frac{\alpha}{2}-\frac{2}{\alpha})\sqrt{1-\mu^2}$ shows that the observer-velocity perturbation dominates once $\alpha\gg1$; since $e^{\pm i\varphi}\sqrt{1-\mu^2}\propto Y_{1,\pm1}(\theta,\varphi)$ is a pure $\ell=1$ dipole, Eq.~\eqref{eq:masterE} reduces to the quadratically growing dipolar correlation
\begin{equation}\label{eq:IR_h1}
    2\Gamma_{+1}^{\mathrm E}(\zeta;\alpha)
    =\frac14\left(\alpha-\frac{4}{\alpha}\right)^2\cos\zeta+\mathcal O(\alpha^{-3}).
\end{equation}
For tensor and higher-helicity modes ($|h|=n\ge2$), the static-observer velocity perturbation vanishes and rapid Bessel oscillations suppress the sky integral. Evaluating Eq.~\eqref{eq:zero_lag_xintegral} via Mellin--Barnes contour integration reveals that parity in the $x$-expansion eliminates all even negative powers of $\alpha$, so that the $\alpha^{-6}$ term vanishes identically, yielding the asymptotic expansion
\begin{equation}\label{eq:IR_hn}
    2\Gamma_{+n}^{\mathrm E}(0;\alpha)
    =\frac{30}{\alpha^5}-\frac{105(4n^2-1)}{4\alpha^7}+\mathcal O(\alpha^{-9}),
\end{equation}
valid when the continuous-spin parameter satisfies $\alpha\gg n$. For the tensor sector ($n=2$), Eq.~\eqref{eq:IR_hn} gives $D_T(\alpha)\simeq 60/\alpha^5$ at $\alpha\gg1$. At $\alpha\gg1$, the integral weight of $\int_{-1}^1 A_n(\mu;\alpha)^2\,\dd\mu$ is distributed across the backward hemisphere $\mu<0$ ($x\lesssim 1$; e.g., accounting for ${95.3}\%$ of the $n=2$ integral at $\alpha=9$), where $J_n(\alpha x)$ oscillates rapidly with an effective angular wavenumber $\sim\alpha/2$ and shifts the dominant multipoles from $\ell=|h|$ out to $\ell\sim\mathcal O(\alpha)$. Table~\ref{tab:scaling_summary} summarizes the multipole selection rules, observer-velocity perturbations, and asymptotic scalings across all helicity sectors.

\subsection{Numerical behavior across helicities and continuous-spin scales}
\label{sec:orf_numerical}

\begin{figure*}[!t]
    \centering
    \includegraphics[width=0.86\textwidth,trim=0 7bp 0 6bp,clip]{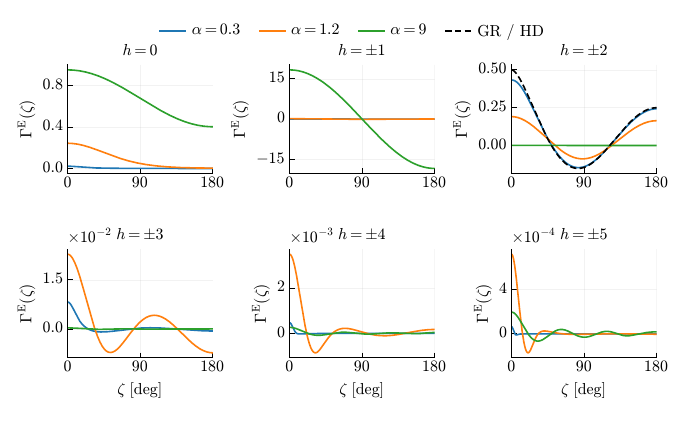}
    \caption{Earth-term overlap reduction functions $\Gamma^{\mathrm E}(\zeta)$ across continuous-spin regimes $\alpha=\rho_g/\omega=0.3$ (blue), $1.2$ (orange), and $9$ (green) for helicities $h=0,\pm1,\pm2,\pm3,\pm4,\pm5$. Here $\Gamma_0^{\mathrm E}(\zeta;\alpha)$ is shown for $h=0$ and the unpolarized sum $\Gamma_{+n}^{\mathrm E}(\zeta;\alpha)+\Gamma_{-n}^{\mathrm E}(\zeta;\alpha)$ is shown for $n\ge1$, with the GR Hellings--Downs curve $\Gamma_{\rm HD}(\zeta)$ (black dashed) overlaid in the $h=\pm2$ panel. Vertical scales differ among panels to accommodate the wide dynamic range across $\alpha$.}
    \label{fig:low}
\end{figure*}

\begin{figure*}[!t]
    \centering
    \includegraphics[width=0.84\textwidth]{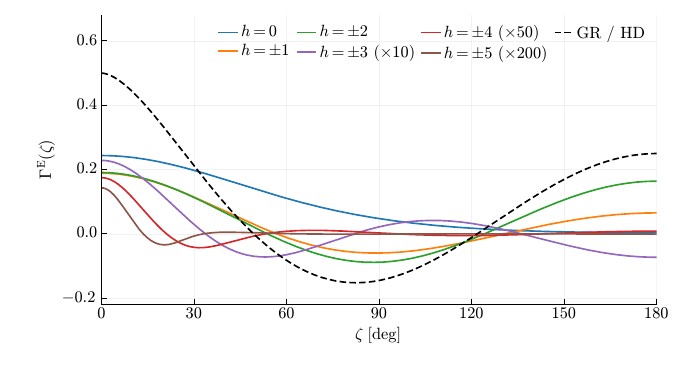}
    \caption{Earth-term angular correlation functions $\Gamma^{\mathrm E}(\zeta)$ for different helicities at $\alpha=1.2$ (defined as in Fig.~\ref{fig:low}). The black dashed curve is the Hellings--Downs curve of general relativity. For visibility, the curves for $h=\pm3,\pm4,\pm5$ are multiplied by $10$, $50$, and $200$, respectively, as indicated in the legend.}
    \label{fig:helicity}
\end{figure*}

\begin{figure*}[!t]
    \centering
    \includegraphics[width=0.84\textwidth,trim=0 7bp 0 3bp,clip]{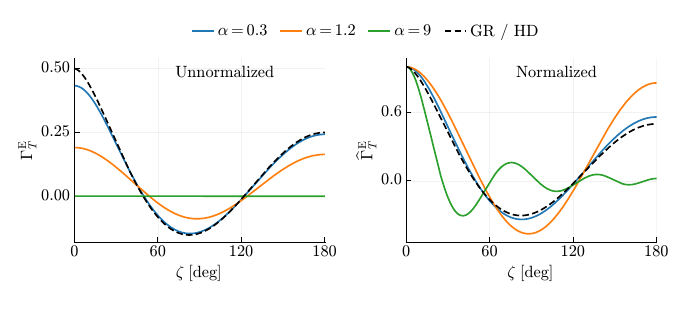}
    \caption{Amplitude suppression and angular-shape deformation of the tensor correlation $\Gamma_T^{\mathrm E}(\zeta;\alpha)=2\Gamma_{+2}^{\mathrm E}(\zeta;\alpha)$ for $\alpha=0.3,1.2,9$ compared with GR (black dashed). The left panel shows the unnormalized Earth-term correlation $\Gamma_T^{\mathrm E}(\zeta;\alpha)$, while the right panel displays the normalized correlation $\widehat\Gamma_T^{\mathrm E}(\zeta;\alpha)\equiv\Gamma_T^{\mathrm E}(\zeta;\alpha)/\Gamma_T^{\mathrm E}(0;\alpha)$ alongside $\widehat\Gamma_{\rm HD}(\zeta)=2\Gamma_{\rm HD}(\zeta)$.}
    \label{fig:tensor_norm}
\end{figure*}

Figures~\ref{fig:low}--\ref{fig:tensor_norm} display the numerical evaluation of Eq.~\eqref{eq:masterE} for $\Gamma_0^{\mathrm E}$ ($h=0$) and the unpolarized pairs $\Gamma_{+n}^{\mathrm E}+\Gamma_{-n}^{\mathrm E}=2\Gamma_{+n}^{\mathrm E}$ ($1\le n\le5$) at three representative scale ratios, $\alpha=0.3$, $1.2$, and $9$, corresponding to the ultraviolet ($\alpha<1$), intermediate ($\alpha\sim1$), and deep infrared ($\alpha\gg1$) regimes. Figure~\ref{fig:low} directly illustrates the three helicity regimes established analytically in Sec.~\ref{sec:asymptotics}. At $\alpha=0.3$ (blue curves), little-group mixing is confined to the narrow collinear region $1-\mu\lesssim\alpha^2$, so the tensor correlation closely tracks $\Gamma_{\rm HD}(\zeta)$ with mild zero-lag damping ($\Gamma_T^{\mathrm E}(0;0.3)\simeq{0.433}$ versus $0.5$), whereas all non-tensor correlations remain suppressed in accordance with Eqs.~\eqref{eq:UV_h0}--\eqref{eq:UV_hn}. At $\alpha=9$ (green curves), the three infrared behaviors are clearly separated. The scalar correlation ($h=0$) approaches the positive monopole-plus-dipole limit~\eqref{eq:IR_h0}, reaching $\Gamma_0^{\mathrm E}\simeq{0.95}$ at $\zeta=0$ and ${0.40}$ at $\zeta=\pi$. The vector correlation ($h=\pm1$) is dominated by the observer-velocity dipole term~\eqref{eq:IR_h1}, reaching $\pm{18.3}$ at $\zeta=0,\pi$, far above the $\alpha=0.3$ and $1.2$ curves of zero-lag amplitudes ${0.068}$ and ${0.189}$. All $|h|\ge2$ correlations undergo the $\alpha^{-5}$ suppression of Eq.~\eqref{eq:IR_hn}, dropping to ${4.3\times10^{-4}}$ for $h=\pm2$ and ${(2.0\text{--}3.5)\times10^{-4}}$ for $|h|=3,4,5$, so that the green curves lie nearly flat along zero on the vertical scale of the $h=\pm2,\pm3$ panels. Because $|h|\ge3$ correlations vanish in both the $\alpha\to0$ and $\alpha\to\infty$ limits, their amplitude varies non-monotonically with $\alpha$ and peaks at $\alpha\sim\mathcal O(|h|)$, which explains why the intermediate $\alpha=1.2$ (orange) curves have larger peak magnitudes than those for $\alpha=0.3$ and $\alpha=9$ in the $|h|=3,4,5$ panels.

Figure~\ref{fig:helicity} compares the angular correlation functions for $h=0$ through $h=\pm5$ at $\alpha=1.2$, rescaling $h=\pm3,\pm4,\pm5$ by factors of $10$, $50$, and $200$. Under the normalization of Eq.~\eqref{eq:Fgeneral}, prior to rescaling, the lower-helicity modes ($h=0,\pm1,\pm2$) attain comparable zero-lag amplitudes (${0.244}$, ${0.189}$, and ${0.191}$), whereas the higher-helicity amplitudes drop rapidly (${2.3\times10^{-2}}$, ${3.5\times10^{-3}}$, and ${7.2\times10^{-4}}$) because $J_n(z)$ is strongly suppressed for $z\lesssim n$. On this scale, the multipole structure of each helicity is directly visible. Because $J_0(z)-1+z^2/4>0$ for all $z>0$, the polar profile satisfies $A_0(\mu;\alpha)\ge 0$ across $[-1,1]$ (and is strictly positive for $\mu>-1$), ensuring that the scalar Earth-term overlap reduction function $\Gamma_0^{\mathrm E}(\zeta;\alpha)$ remains strictly positive at all separations. At $\alpha=1.2$, partial cancellation between the photon-momentum and observer-velocity dipole terms in Eq.~\eqref{eq:A1} suppresses the $\ell=1$ harmonic of $2\Gamma_{+1}^{\mathrm E}$, producing a curve with two zero crossings similar to $\Gamma_T^{\mathrm E}$. The pure dipole at $\alpha\gg1$ is shown in Fig.~\ref{fig:low}. For $|h|=2,3,4,5$, the correlation curves exhibit ${2},{3},{4}$, and ${5}$ zero crossings on $\zeta\in(0,\pi)$, with the first zero crossing shifting progressively toward smaller angles ($\zeta\simeq{54}^\circ,{34}^\circ,{21}^\circ,{14}^\circ$). The multipole selection rule $\ell\ge|h|$ in Eq.~\eqref{eq:masterE} makes the angular correlation function orthogonal to all Legendre polynomials of degree below $|h|$, requiring at least $|h|$ sign changes over $0<\zeta<\pi$. The numerical curves at $\alpha=1.2$ attain this lower bound.

Because our NG15 data analysis in Sec.~\ref{sec:data} focuses on the tensor sector ($h=\pm2$), Fig.~\ref{fig:tensor_norm} separates overall amplitude damping from purely geometric distortion in $\Gamma_T^{\mathrm E}(\zeta;\alpha)$ as $\alpha=\rho_g/(2\pi\hbar f)$ varies with frequency (with explicit factors of $\hbar$ and $c$ restored hereafter). While the left panel shows the rapid monotonic drop in overall amplitude and zero-lag correlation $\Gamma_T^{\mathrm E}(0;\alpha)$, the right panel isolates angular shape via the normalized correlation
\begin{equation}\label{eq:tensor_shape_norm}
    \widehat\Gamma_T^{\mathrm E}(\zeta;\alpha)
    \equiv\frac{\Gamma_T^{\mathrm E}(\zeta;\alpha)}
    {\Gamma_T^{\mathrm E}(0;\alpha)},
\end{equation}
compared with $\widehat\Gamma_{\rm HD}(\zeta)=2\Gamma_{\rm HD}(\zeta)$. Moderate values ($\alpha=0.3,1.2$) shift the zero-crossing angles
and make the negative minimum deeper than that of the HD curve,
whereas $\alpha=9$ produces multiple oscillations at intermediate
angular separations. For $\alpha\gg1$, rapid Bessel oscillations across the backward hemisphere $\mu<0$ (Sec.~\ref{sec:asymptotics}) shift the dominant Legendre weights in $\Gamma_T^{\mathrm E}(0;\alpha)$ from the GR quadrupole $\ell=2$, which accounts for $62.5\%$ of $\Gamma_T^{\mathrm E}(0;0)$, to higher multipoles $\ell\sim\mathcal O(\alpha)$, with $\ell=4,5,6$ accounting for ${85.4}\%$ of $\Gamma_T^{\mathrm E}(0;9)$. A continuous-spin tensor background therefore suppresses low-frequency timing power while distorting the angular dependence of inter-pulsar correlations toward higher multipoles.

\section{Analysis of the NANOGrav 15-year data set}
\label{sec:data}

\subsection{Bayesian likelihood and MCMC framework}
\label{sec:data_methods}

We now test the continuous-spin tensor model against the public timing data from the 67 millisecond pulsars in the NG15 SGWB analysis~\cite{NANOGrav:2023hde,NANOGrav:2023ctt,NANOGrav:2023gor}, assuming an unpolarized $h=\pm2$ background ($S_{+2}=S_{-2}$, $S_{h\neq\pm2}=0$) in the solar-system barycentric (SSB) frame; because the peculiar velocities of the SSB and Galactic pulsars relative to the cosmic rest frame are only $\mathcal O(10^{-4}\text{--}10^{-3})c$, giving $\alpha(v/c)\lesssim {10^{-2}}$, boost-induced helicity mixing is negligible. Following the standard NG15 protocol~\cite{NANOGrav:2023gor,NANOGrav:2023icp}, we adopt the JPL DE440 solar-system ephemeris~\cite{Park2021DE440} to refer pulse arrival times to the SSB, the TT(BIPM2019) time standard, and the observing span $T\simeq16.03\,\mathrm{yr}$, expanding stochastic processes on harmonic frequencies $f_k=k/T$ ($k\in\mathbb Z^+$) with the first $14$ frequencies modeling the common gravitational-wave background and the first $30$ frequencies modeling intrinsic per-pulsar red noise.

Within the linear timing-model and Gaussian-process framework~\cite{Lentati:2012xb,vanHaasteren:2012hj}, the concatenated timing-residual vector $\boldsymbol r$ across all 67 pulsars is modeled as
\begin{equation}\label{eq:data_residual_model}
    \boldsymbol r=M\boldsymbol\epsilon
    +F_{\rm RN}\boldsymbol a_{\rm RN}
    +F_{\rm GW}\boldsymbol a_{\rm GW}
    +\boldsymbol n,
\end{equation}
where $M$ is the timing-model design matrix with linear parameter offsets $\boldsymbol\epsilon$, $F_{\rm RN}$ and $F_{\rm GW}$ are the Fourier design matrices for intrinsic red noise and the common background with zero-mean Gaussian coefficients $\boldsymbol a_{\rm RN}$ and $\boldsymbol a_{\rm GW}$, and $\boldsymbol n$ denotes white noise of covariance $N$. White-noise parameters (EFAC, EQUAD, and ECORR) are fixed to their NG15 single-pulsar noise-run values~\cite{NANOGrav:2023ctt,NANOGrav:2023icp}, while all red-noise and common-background hyperparameters are inferred jointly.

In the short-wavelength regime, combining the coherent Earth-term correlation with the equal diagonal pulsar-term contribution yields the $67\times67$ full tensor correlation matrix at frequency $f_k$:
\begin{equation}\label{eq:data_full_orf}
    \Gamma_{T,ab}(\rho_g,f_k)
    =\Gamma_T^{\mathrm E}(\zeta_{ab};\alpha_k)
    +\delta_{ab}\Gamma_T^{\mathrm E}(0;\alpha_k),
\end{equation}
where $a,b\in\{1,\ldots,67\}$ index pulsars separated by angle $\zeta_{ab}$, $\delta_{ab}$ is the Kronecker delta, and $\alpha_k\equiv\rho_g/(2\pi\hbar f_k)$. By Parseval's identity, the diagonal autocorrelation entries ($a=b$) of Eq.~\eqref{eq:data_full_orf} equal
\begin{equation}\label{eq:data_auto}
\begin{aligned}
    D_T(\alpha)&\equiv\Gamma_{T,aa}(\rho_g,f)=2\Gamma_T^{\mathrm E}(0;\alpha)\\
    &=\frac32\int_{-1}^{1}\dd\mu\,A_2(\mu;\alpha)^2,
\end{aligned}
\end{equation}
approaching $D_T\to1$ and $\Gamma_{T,ab}\to\Gamma_{\rm HD}(\zeta_{ab})+\frac12\delta_{ab}$ as $\alpha\to0$~\cite{Hellings:1983fr,Anholm:2008wy}.

We model the common-background one-sided timing-residual power spectral density $S_r(f)$, which is related to $S_{\pm2}(f)$ in Eq.~\eqref{eq:cross_final} by $S_r(f)=\frac{4}{3(2\pi f)^2}\allowbreak[S_{+2}(f)+S_{-2}(f)]$, as a power law referenced to $f_{\rm yr}=1\,\mathrm{yr}^{-1}$~\cite{Anholm:2008wy,Romano:2016dpx,NANOGrav:2023gor},
\begin{equation}\label{eq:data_psd}
    S_r(f)=\frac{A_{\rm GW}^2}{12\pi^2}f_{\rm yr}^{-3}
    \left(\frac{f}{f_{\rm yr}}\right)^{-\gamma_{\rm GW}},
\end{equation}
with dimensionless strain amplitude $A_{\rm GW}$ and spectral index $\gamma_{\rm GW}$, alongside an independent power law $(A_{{\rm RN},a},\gamma_{{\rm RN},a})$ for each pulsar's intrinsic red noise~\cite{NANOGrav:2023ctt}. In the Fourier-basis Gaussian-process representation~\cite{Lentati:2012xb}, cosine and sine basis functions at frequency $f_k$ share the $67\times67$ inter-pulsar covariance blocks
\begin{equation}\label{eq:data_phi}
    (\Phi_{{\rm GW},k})_{ab}
    =\frac{S_r(f_k)}{T}
    \Gamma_{T,ab}(\rho_g,f_k),
\end{equation}
assembled across $k=1,\ldots,14$ into $\Phi_{\rm GW}$. Integrating over $\boldsymbol a_{\rm RN}$ (with block-diagonal covariance $\Phi_{\rm RN}$), $\boldsymbol a_{\rm GW}$, and $\boldsymbol n$ yields the total residual covariance~\cite{Chamberlin:2014ria}
\begin{equation}\label{eq:data_total_covariance}
    C=N+F_{\rm RN}\Phi_{\rm RN}F_{\rm RN}^T +F_{\rm GW}\Phi_{\rm GW}F_{\rm GW}^T.
\end{equation}
Analytic marginalization over the deterministic timing offsets $\boldsymbol\epsilon$~\cite{vanHaasteren:2012hj,Lentati:2012xb,vanHaasteren:2014qva} projects $\boldsymbol r$ onto the orthogonal complement of the column space of $M$ via an orthonormal basis matrix $G$ ($\boldsymbol r_G=G^T\boldsymbol r$, $C_G=G^TCG$), giving the marginalized Gaussian log-likelihood for the $137$-dimensional hyperparameter vector $\boldsymbol\Theta$ ($134$ pulsar red-noise parameters plus $\log_{10}A_{\rm GW}$, $\gamma_{\rm GW}$, and $\rho_g$),
\begin{equation}\label{eq:data_likelihood}
    \ln\mathcal L(\boldsymbol r_G\mid\boldsymbol\Theta) =-\frac12\Big[\boldsymbol r_G^TC_G^{-1}\boldsymbol r_G+\ln\det C_G +n_G\ln(2\pi)\Big],
\end{equation}
where $n_G=\dim(\boldsymbol r_G)$. Equation~\eqref{eq:data_likelihood} uses a $G$-matrix projection onto the null space of $M^T$. In practice, we evaluate the equivalent marginalized likelihood using the Woodbury matrix identity, as implemented in \texttt{enterprise}~\cite{Enterprise2020,vanHaasteren:2014qva,NANOGrav:2023icp}.

\begin{table}[!htbp]
\caption{Sampled model parameters and their prior distributions ($a=1,\ldots,67$ indexes pulsars).}
\label{tab:priors}
\centering
\setlength{\tabcolsep}{10pt}
\renewcommand{\arraystretch}{1.15}
\begin{tabular*}{\columnwidth}{@{\extracolsep{\fill}}ll@{}}
\toprule
Parameter & Prior distribution\\
\midrule
$\log_{10}A_{{\rm RN},a}$ & $\mathcal U(-20,-11)$\\
$\gamma_{{\rm RN},a}$ & $\mathcal U(0,7)$\\
$\log_{10}A_{\rm GW}$ & $\mathcal U(-18,-11)$\\
$\gamma_{\rm GW}$ & $\mathcal U(0,7)$\\
$\rho_g\,[\mathrm{eV}]$ & $\mathcal U(0,10^{-20})$\\
$\log_{10}(\rho_g/\mathrm{eV})$ & $\mathcal U(-30,-20)$\\
\bottomrule
\end{tabular*}
\end{table}

Table~\ref{tab:priors} lists the prior distributions $\pi(\boldsymbol\Theta)$, where $\mathcal U(a,b)$ denotes a uniform distribution on $[a,b]$ and the red-noise and common-spectrum priors follow standard NG15 conventions~\cite{NANOGrav:2023gor,NANOGrav:2023icp}. For $\rho_g$, we consider both uniform and log-uniform priors. We sample the $137$-dimensional posterior via Markov chain Monte Carlo (MCMC)~\cite{Hastings:1970aa,vanHaasteren:2008yh} using \texttt{PTMCMCSampler}~\cite{PTMCMCSampler2017,NANOGrav:2023icp}. Convergence is assessed using the rank-normalized split-$\widehat R$ and effective sample size (ESS) statistics~\cite{Vehtari2021}, with thresholds of $\widehat R\le1.01$, bulk $\mathrm{ESS}\ge400$, and tail $\mathrm{ESS}\ge400$. 

\begin{figure*}[!tp]
    \centering
    \includegraphics[width=0.82\textwidth]{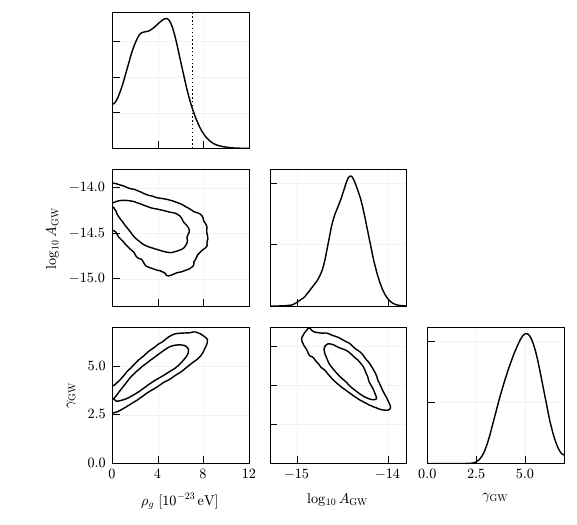}
    \caption{Marginalized one- and two-dimensional posterior distributions of $\rho_g$,  $\log_{10}A_{\rm GW}$, $\gamma_{\rm GW}$ under the uniform prior on $\rho_g$. The vertical dotted line in the upper-left panel marks the $95\%$ credible upper limit $\rho^{\mathcal{U}}_{95}={7.03\times10^{-23}}\,\mathrm{eV}$; inner and outer contours in the two-dimensional panels enclose $68\%$ and $95\%$ credible regions, respectively.}
    \label{fig:joint_corner}
\end{figure*}

\begin{figure*}[!tp]
    \centering
    \includegraphics[width=0.82\textwidth]{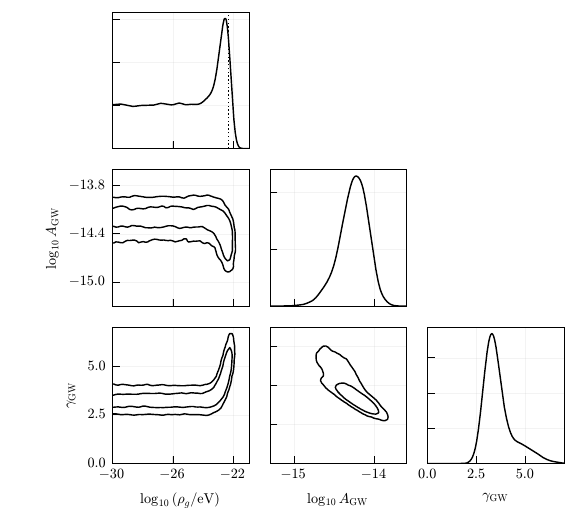}
    \caption{Same with Fig.~\ref{fig:joint_corner}, but under the log-uniform prior on $\rho_g$. The vertical dotted line in the upper-left panel marks the $95\%$ upper limit, $\rho^{\log\mathcal{U}}_{95}=4.79\times10^{-23}\,\mathrm{eV}$.}
    \label{fig:logprior}
\end{figure*}

\subsection{Posterior constraints, spectral degeneracies, and comparison with massive gravity}
\label{sec:data_results}

Figure~\ref{fig:joint_corner} shows the marginalized one- and two-\hspace{0pt}dimensional posteriors of $\rho_g$, $\log_{10}A_{\rm GW}$, and $\gamma_{\rm GW}$. The marginalized density $p(\rho_g)$ remains compatible with the GR limit $\rho_g=0$, concentrating below $\sim8\times10^{-23}\,\mathrm{eV}$ with a broad maximum near ${4.7\times10^{-23}}\,\mathrm{eV}$ (median ${3.86\times10^{-23}}\,\mathrm{eV}$). The $95\%$ credible upper limit $\rho^{\mathcal{U}}_{95}$ is
\begin{equation}\label{eq:rho95_result}
    \rho^{\mathcal{U}}_{95}={7.03\times10^{-23}}\,\mathrm{eV},
\end{equation}
which corresponds to $\hbar c/\rho_g\gtrsim0.091\,\mathrm{pc}$. The marginalized common-background spectral parameters are constrained to
$\log_{10} A_{\mathrm{GW}} = -14.43^{+0.18}_{-0.20}$
and $\gamma_{\mathrm{GW}} = 4.89^{+0.85}_{-0.96}$
(68\% credible intervals). Across the 14 NG15 Fourier bins (from $f_1=1/T\simeq1.98\,\mathrm{nHz}$ to $f_{14}\simeq27.68\,\mathrm{nHz}$), setting $\rho_g=\rho^\mathcal{U}_{95}$ spans $\alpha_1\simeq{8.60}$ down to $\alpha_{14}=\alpha_1/14\simeq{0.61}$, showing that the joint likelihood is sensitive to all frequency bins.

The joint posteriors in Fig.~\ref{fig:joint_corner} show a strong positive correlation between $\rho_g$ and $\gamma_{\rm GW}$, alongside a moderate anticorrelation between $\rho_g$ and $\log_{10}A_{\rm GW}$. In the standard GR analysis, NG15 favors a relatively shallow spectral index $\gamma_{\rm GW}\simeq3.2$~\cite{NANOGrav:2023gor}, below the canonical SMBHB prediction $\gamma_{\rm GW}=13/3$~\cite{Phinney:2001di,NANOGrav:2023hfp}. In continuous-spin gravity, increasing $\rho_g$ suppresses the low-frequency timing response through $D_T(\alpha_k)$, while a larger $\gamma_{\rm GW}$ enhances the underlying low-frequency background power, partially compensating for this suppression. This explains the positive correlation and allows steeper background spectra, including $\gamma_{\rm GW}=13/3$, for $\rho_g\sim(3\text{--}5)\times10^{-23}\,\mathrm{eV}$. Since the best-constrained frequencies ($3\text{--}10\,\mathrm{nHz}$) lie below the reference frequency $f_{\rm yr}\simeq31.7\,\mathrm{nHz}$, increasing $\gamma_{\rm GW}$ also tends to lower the inferred amplitude $A_{\rm GW}$ at $f_{\rm yr}$, contributing to the observed anticorrelation.

The posterior distributions under the log-uniform prior on $\rho_g$ is plotted in Figure~\ref{fig:logprior}. The $95\%$ upper limit is
\begin{equation}\label{eq:rho95_log_result}
    \rho^{\log \mathcal{U}}_{95}=4.79\times10^{-23}\,\mathrm{eV},
\end{equation}
which corresponds to $\hbar c/\rho_g\gtrsim0.135\,\mathrm{pc}$. The log-uniform prior assigns greater weight to smaller continuous-spin scales, leading to a smaller upper limit than the uniform prior.

It is instructive to contrast this continuous-spin bound with nanohertz constraints on a massive graviton of mass $m_g$, which also induces frequency-dependent PTA correlations~\cite{Lee:2010cg,Liang:2021bct,Bernardo:2023mxc,Liang:2024mex,Wang:2023div}. Freely propagating massive gravitons obey the dispersion relation and propagation threshold
\begin{equation}\label{eq:data_massive_dispersion}
    \omega^2=c^2|\boldsymbol{k}|^2+
    \left(\frac{m_gc^2}{\hbar}\right)^2,
    \qquad
    f\ge f_g\equiv\frac{m_gc^2}{2\pi\hbar},
\end{equation}
below which oscillatory plane waves become evanescent; requiring the lowest NG15 frequency bin $f_1\simeq1.98\,\mathrm{nHz}$ to propagate ($f_g\le f_1$) immediately yields $m_g<8.2\times10^{-24}\,\mathrm{eV}/c^2$~\cite{Wu:2023rib,Wang:2023div}. In contrast, continuous-spin gravitational waves remain strictly massless and luminal ($\omega=c|\boldsymbol k|$) at all frequencies~\cite{Schuster:2014hca,Kundu:2025fsd}, so no sharp dispersion cutoff exists. Instead, continuous-spin gravity suppresses timing power through a smooth, frequency-dependent damping factor $D_T(\alpha)$, which decreases monotonically. At the $95\%$ upper limit $\rho_g=\rho^\mathcal{U}_{95}$ ($\alpha_1\simeq{8.60}$), timing power in the lowest four frequency bins (where $\alpha_k>{2}$) is suppressed by $>{85}\%$. In a joint Bayesian fit, this low-frequency deficit is partially compensated by steepening the common-background spectral index $\gamma_{\rm GW}$, until the combination of residual power at intermediate frequencies and higher-multipole spatial correlation distortions ($\ell\sim\mathcal O(\alpha)$) excludes larger continuous-spin scales. Together, the absence of a hard dispersion cutoff and this spectral degeneracy allow $\rho^\mathcal{U}_{95}\simeq{7.03\times10^{-23}}\,\mathrm{eV}$ to exceed the massive-gravity threshold energy $2\pi\hbar f_1\simeq8.2\times10^{-24}\,\mathrm{eV}$ by nearly an order of magnitude.

\section{Conclusions}\label{sec:concl}

In this work, we derived the one-way pulsar timing redshift response and the overlap reduction function $\Gamma_h^{\mathrm E}(\zeta;\alpha)$ of an isotropic stochastic background in linearized continuous-spin gravity for all integer helicities $h\in\mathbb Z$. From the Legendre expansion~\eqref{eq:masterE}, we obtained closed-form ultraviolet and infrared scaling laws governed by $\alpha=\rho_g/(\hbar\omega)$. We find that while photon propagation is modulated across every helicity, rotational symmetry in the observer rest frame restricts the velocity perturbation of a static observer to the vector modes ($h=\pm1$). For $\alpha\ll1$, collinear helicity mixing within $1-\mu\lesssim\alpha^2$ lifts all $|h|\ge4$ correlations to $\mathcal O(\alpha^2)$ and produces $\alpha^2\ln(1/\alpha)$ enhancements for $|h|=1,2,3$. For $\alpha\gg1$, the vector correlation grows quadratically as an $\ell=1$ dipole, the scalar correlation ($h=0$) approaches a constant monopole-plus-dipole form, and all $|h|\ge2$ correlations decay as $\alpha^{-5}$ while their dominant multipoles shift to $\ell\sim\mathcal O(\alpha)$. In the limit $\alpha\to0$, all non-tensor responses vanish and the tensor sector smoothly recovers the GR Hellings--Downs curve.

In a Bayesian analysis of NG15 data, fitting the frequency-dependent tensor correlation matrix $\Gamma_{T,ab}(\rho_g,f_k)$ under a uniform prior on $\rho_g$ yields the $95\%$ credible upper limit $\rho^\mathcal{U}_{95}<{7.03\times10^{-23}}\,\mathrm{eV}$. With the log-uniform prior, the $95\%$ upper limit is $\rho^{\log \mathcal{U}}_{95}<4.79\times10^{-23}\,\mathrm{eV}$. Continuous-spin damping suppresses the lowest Fourier bins ($\alpha_k\propto f_k^{-1}$), producing a low-frequency turnover in the timing spectrum that induces a strong positive correlation between $\rho_g$ and the common-background spectral index $\gamma_{\rm GW}$, which brings steeper background spectra, including the canonical $\gamma_{\rm GW}=13/3$ of circular SMBHBs, into closer agreement with the NG15 data.

Although a strong-field calculation of binary radiation in continuous-spin gravity remains an open problem, helicity correspondence~\cite{Schuster:2013vpr,Schuster:2023xqa,Kundu:2025fsd} suppresses non-tensor emission from non-relativistic binaries when $\alpha(v/c)\ll 1$. If relativistic or primordial sources generate appreciable scalar or vector radiation, the spatial correlation functions derived here, in particular the monopole-dominated $h=0$ correlation and the dipolar $h=\pm1$ correlation at large $\alpha$, allow direct searches for non-tensor continuous-spin polarizations in PTA data. Because continuous-spin effects depend on $\rho_g/(\hbar\omega)$, nanohertz observations probe values of $\rho_g$ roughly five orders of magnitude smaller than millihertz space-based interferometers and ten orders of magnitude smaller than ground-based ($\sim100\,\mathrm{Hz}$) detectors~\cite{Kundu:2025fsd}. Longer-baseline observations with the International Pulsar Timing Array and the Square Kilometre Array will improve sensitivity at the lowest frequencies and across angular multipoles, helping to separate finite-$\rho_g$ angular distortion from the common-background spectral slope $\gamma_{\rm GW}$.

\paragraph{Note added.}
While this manuscript was being finalized, an independent study~\cite{Ambasht:2026anp} appeared, analyzing continuous-spin tensor gravitons in the NG15 data set and quoting a $95\%$ upper limit $\rho_g<1.23\times10^{-23}\,\mathrm{eV}$. Our analytical tensor response agrees exactly with that of Ref.~\cite{Ambasht:2026anp}, while our theoretical derivation additionally covers all non-tensor helicities ($h=0,\pm1$, and $|h|\ge3$), closed-form asymptotic scaling laws, and the trace correction to the massive-particle worldline kernel in Eq.~\eqref{eq:jkernel}. This correction also ensures the proper GR response limit, as only the $h=\pm2$ modes retain the standard tensor response when $\rho_g\to0$, while the responses of all other helicities vanish. 

The numerical difference between our uniform-prior upper limit and the limit reported in that work arises primarily from the choice of prior. Reference~\cite{Ambasht:2026anp} adopts a log-uniform prior extending several decades below $10^{-23}\,\mathrm{eV}$, where the likelihood approaches its nonzero GR limit. Lowering the cutoff assigns more prior probability to the region $\rho_g\ll10^{-23}\,\mathrm{eV}$. In contrast, our uniform prior on $\rho_g$ is independent of any nonzero lower cutoff and directly probes the scale at which the NG15 likelihood begins to decrease. 

The results difference between our log-uniform-prior limit and the limit reported in that work may mainly arising from the treatment of pulsar autocorrelations(and partially from the slight difference in prior boundaries). Our analysis retains the full frequency-dependent autocorrelation $D_T(\alpha_k)=2\Gamma_T^{\mathrm E}(0;\alpha_k)$, which is expected to induce a strong positive correlation between $\rho_g$ and $\gamma_{GW}$. Furthermore, inspecting the publicly released implementation and MCMC chains of Ref.~\cite{Ambasht:2026anp} indicates that their likelihood evaluation adopts the standard convention ($\Gamma_{aa}=1$) commonly hardcoded in some PTA software interfaces, whereas our Eq.~(67) retains the full frequency-dependent autocorrelation $D_T(\alpha_k) = 2\Gamma_T^E(0;\alpha_k)$. Under a comparable log-uniform prior, our full-covariance analysis yields $\rho^{\log \mathcal{U}}_g < 4.79 \times 10^{-23}\,\mathrm{eV}$ (Fig.~5). 

\section{Acknowledgements}
This work is supported by the 2115 Talent Development Program of China Agricultural University, and the High-performance Computing Platform of China Agricultural University. The Large Language Model was used to check some calculations and to assist with editing the manuscript.


\bibliographystyle{spphys}
\bibliography{references}
\end{document}